\documentclass[twocolumn]{aastex631}

\usepackage{amsmath}	
\usepackage{amssymb}	
\usepackage{bm}		
\usepackage{xcolor}

\newcommand*{\Kp}{\ensuremath{\text{K}_\text{p}}}
\newcommand*{\Vsys}{\ensuremath{\text{V}_\text{sys}}}
\newcommand*{\pipeline}{ExoRES }

\shorttitle{A First Look at CRIRES+}
\shortauthors{Holmberg \& Madhusudhan}

\begin{document}

\title{A First Look at CRIRES+: Performance Assessment and Exoplanet Spectroscopy} 

\author{M\aa ns Holmberg}
\affiliation{Institute of Astronomy, University of Cambridge, Madingley Road, Cambridge, CB3 0HA, UK}
\author{Nikku Madhusudhan}
\affiliation{Institute of Astronomy, University of Cambridge, Madingley Road, Cambridge, CB3 0HA, UK}

 \correspondingauthor{Nikku Madhusudhan, M\aa ns Holmberg}
 \email{nmadhu@ast.cam.ac.uk, mlh58@cam.ac.uk}

\begin{abstract}
High-resolution spectroscopy has proven to be a powerful avenue for atmospheric remote sensing of exoplanets. Recently, ESO commissioned the CRIRES+ high-resolution infrared spectrograph at VLT. CRIRES+ is a cross-dispersed spectrograph with high throughput and wide wavelength coverage across the near-infrared (0.95-5.3 $\mu$m), designed to be particularly suited for atmospheric characterisation of exoplanets. In this work, we report early insights into the performance of CRIRES+ for exoplanet spectroscopy and conduct a detailed assessment of the data reduction procedure. Because of the novelty of the instrument, we perform two independent data reduction strategies, using the official CR2RES pipeline and our new custom-built \pipeline pipeline. Using science verification observations we find that the spectral resolving power of CRIRES+ can reach $R \gtrsim 100,000$ for optimal observing conditions. Similarly, we find the signal-to-noise ratio (S/N) to  be consistent with expected and empirical estimates for the observations considered. As a case study, we perform the first application of CRIRES+ to the atmospheric characterisation of an exoplanet - the ultra-hot Jupiter MASCARA-1 b. We detect CO and H$_2$O in the atmosphere of MASCARA-1 b at  a S/N of 12.9 and 5.3, respectively, and a temperature inversion revealed through the CO and H$_2$O emission lines, the first for an exoplanet. We find a combined S/N of 13.8 for CO and H$_2$O together, with a preference for lower H$_2$O abundance compared to CO. Our findings demonstrate the scientific potential of CRIRES+ and highlight the excellent opportunity for high-resolution atmospheric spectroscopy of diverse exoplanets.
\end{abstract}


\section{Introduction} \label{sec:intro}

High-resolution Doppler spectroscopy has been one of the most promising methods for characterising exoplanetary atmospheres in recent years. The method was first demonstrated by the detection of CO in the hot jupiter HD 209458 b \citep{snellen_orbital_2010} using the CRyogenic high-resolution InfraRed Echelle Spectrograph \cite[CRIRES;][]{kaeufl_crires_2004} that was installed on the ESO Very Large Telescope (VLT) until 2014. This method uses the fact that at high spectral resolution, atomic and molecular lines in a planet's spectrum are Doppler shifted due to the orbital motion of the planet, which allows for the removal of quasi-static stellar and telluric lines when observing over time. The high-resolution allows the lines to be individually resolved instead of being blended together. The signatures of chemical species can then be extracted from the noisy residuals by cross-correlating with a model spectrum of the planet's atmosphere, effectively co-adding the signal of all lines within the wavelength range observed, which typically amount to hundreds or even thousands of lines. Prior to the the decommission of old CRIRES, the instrument was used to pioneer the field of high-resolution spectroscopy of exoplanetary atmospheres, leading to detections of key molecules H$_2$O and CO in several hot Jupiters \citep{snellen_orbital_2010, brogi_signature_2012, brogi_detection_2013, brogi_carbon_2014, birkby_detection_2013, de_kok_detection_2013, birkby_discovery_2017}. Because this technique is directly probing the line profile it can also constrain atmospheric dynamics \citep{flowers_high-resolution_2019, ehrenreich_nightside_2020, kesseli_confirmation_2021}, rotation rate \citep{snellen_fast_2014, brogi_rotation_2016} and the temperature profile \citep{schwarz_evidence_2015, brogi_retrieving_2019, gandhi_hydra-h_2019, yan_temperature_2020}. Smaller, 3.6-m class, telescopes have also been used to study exoplanet atmospheres with molecular detections using high-resolution infrared spectrographs with wide spectral coverage \citep[e.g.,][]{brogi_exoplanet_2018,sanchez-lopez_water_2019}, given that the signal-to-noise ratio (S/N) of a planet's spectrum scales as the square root of the number of lines observed \citep{snellen_combining_2015}. Recently, \cite{line_solar_2021} used IGRINS (R = 45,000) on the 8.1m Gemini telescope to obtain high significance molecular detections in WASP-77A b, allowing for constraints on the carbon-to-oxygen (C/O) ratio and metallicity, thus demonstrating the power of combining a large wavelength coverage with a large aperture. 

Recently, CRIRES has been upgraded to a cross-dispersed spectrograph, increasing the instantaneous spectral coverage by up to a factor of ten \citep{ramsay_crires_2014, dorn_crires_2014}. The new instrument, referred to as CRIRES+ or upgraded CRIRES, is ideal for atmospheric characterisation of exoplanets, as it combines the light-gathering power of VLT with a high spectral resolution ($R > 80,000$ for the 0.2" slit) and a large wavelength coverage. CRIRES+ is a high-resolution near-infrared spectrograph (0.95 - 5.3 $\mu$m) that is aided by the Multi Applications Curvature Adaptive Optics system \citep{paufique_macao-crires_2004}. The cross-disperser and echelle gratings are held in a stabilised cryogenic vacuum vessel at 65 K, which also include the detector array, that are further cooled to 35 K. The cross-disperser is made up of six gratings, one for each wavelength band (Y, J, H, K, L and M). Depending on the wavelength setting, the echellogram consists of up to 9 spectral orders. 

To accompany the larger focal plane, three new state-of-the-art 2048$\times$2048 pixel Hawaii 2RG detectors have been installed \citep{ramsay_crires_2014}. A detailed description of the instrument can be found in the CRIRES+ User Manual (P109.4)\footnote{\label{note:crires_user_manual}\href{http://www.eso.org/sci/facilities/paranal/instruments/crires/doc/ESO-254264\_CRIRES\_User\_Manual_P109.4.pdf}{http://www.eso.org/sci/facilities/paranal/instruments/crires/doc/ESO-254264\_CRIRES\_User\_Manual\_P109.4.pdf}}. The improved detector array provides lower noise and an increased pixel response homogeneity, leading to higher quality data. A new calibration unit has been installed to provide improved wavelength calibration capabilities, which include a Fabry-Pérot etalon system and new gas cells. In addition, a new metrology system has been developed to ensure accurate repeatability of the cross-disperser and echelle grating, enabling an absolute wavelength repeatability of $<0.2$ pixels. Together with a $\sim15\%$ increase in overall throughput, CRIRES+ promises a leap in performance for high-resolution infrared spectroscopy\textsuperscript{\ref{note:crires_user_manual}}. 

Two out of the three science goals of CRIRES+ relate to exoplanets \citep{ramsay_crires_2014}: (1) the search for terrestrial planets in the habitable zones around low mass stars, and (2) atmospheric characterisation of exoplanets. Searching for low-mass exoplanets around low-mass stars is particularly interesting as their habitable zones occur close to the host star, resulting in a $\sim$ 1 m s$^{-1}$ radial velocity semi-amplitude for an Earth-mass planet in the habitable zone of a late M dwarf. High-resolution infrared spectroscopy also allows the characterisation of the diverse chemical landscape of exoplanetary atmospheres \citep{snellen_orbital_2010, birkby_exoplanet_2018}. CRIRES+ is well suited for both transmission and emission spectroscopy of exoplanet atmospheres, which in the infrared allow detections of a wide range of molecules such as CO, H$_2$O, CH$_4$, NH$_3$, and HCN \citep{madhusudhan_exoplanetary_2016}.

\begin{table*}
\caption{Overview of the close-in exoplanet observations during the science verification run of CRIRES+. N$_\mathrm{OBS}$ is the number of observed spectra per target, and the exposure time is expressed as NDIT $\times$ DIT, where DIT is the detector integration time and NDIT is the number of detector integrations.}
\movetableright=-0.75in
\scriptsize
\begin{tabular}{lccccccccc}
\hline
Target      & Night & Phase $\phi$      & N$_\mathrm{OBS}$ & Exp. Time    & Obs. Mode &  Slit & AO loop & Wavelength Setting       & Programme ID \\ \hline
WASP-20 b & 2021-09-15  & 0.98 - 0.02 & 75        & 1 $\times$ 180 s  & Nodding & 0.2" & Closed & K2217        & 107.22SX.001 \\
MASCARA-1 b & 2021-09-16 & 0.32 - 0.42 & 107        & 5 $\times$ 30 s  & Staring & 0.2" & Closed& K2166        & 107.22TQ.001 \\
LTT 9779 b   & 2021-09-17 & 0.95 - 0.08 & 76        & 1 $\times$ 120 s & Staring & 0.4" & Closed & K2217 & 107.22TE.001 \\ 
HIP 65A b  & 2021-09-18  & 0.94 - 0.04 & 24        & 1 $\times$ 300 s$^{(\mathrm{a})}$  & Nodding & 0.2" & Open & K2192         & 107.22U7.001 \\\hline
\end{tabular}
\label{tab:obs}
\newline
\footnotesize{\textbf{Notes}. $^{(\mathrm{a})}$Applies to the first 20 exposures, which employed the short gas cell as a simultaneous wavelength calibrator. The last four exposures were taken without a gas cell and with exposure time 1 $\times$ 450 s. \hfill}
\end{table*}

In this work, we aim to obtain early insights into the performance of CRIRES+ and conduct a detailed assessment of the data reduction and application to high-resolution atmospheric spectroscopy of exoplanets. For validation purposes, we developed an independent reduction pipeline ExoRES, alongside the official ESO CR2RES reduction pipeline, which we use throughout this study to check for consistency. Using observations from four exoplanet programs from the science verification we assess the on-sky spectral resolution in a variety of conditions and compare the obtained signal-to-noise ratio (S/N) to the expected S/N from the CRIRES+ Exposure Time Calculator (ETC). 

To further demonstrate the performance and scientific potential of CRIRES+ for atmospheric characterisation of exoplanets, we perform a case study of the Ultra Hot Jupiter (UHJ) MASCARA-1 b observed in thermal emission. MASCARA-1 b (also known as HD 201585 b) is one of the hottest known exoplanets with an equilibrium temperature of 2594 K, orbiting a bright A8 type star ($V=8.3$) with a period of 2.15 days \citep{talens_mascara-1_2017, hooton_spi-ops_2022}. Here we present the first atmospheric characterisation of MASCARA-1 b in search of CO and H$_2$O, as well as the potential existence of a temperature inversion.

 Ultra-hot Jupiters form a class of highly irradiated close-in gas giants with dayside temperatures exceeding 2200 K \citep{arcangeli_h_2018, lothringer_extremely_2018, parmentier_thermal_2018}, which provide excellent laboratories for studying the extremes of atmospheric conditions that are nowhere to be found in the solar system. Their close-in orbits, large radii, and large scale heights make UHJs especially suitable for atmospheric studies. Recently, a large diversity of neutral and ionised atomic metals have been found in UHJs using high-resolution transmission spectroscopy in the optical and near-infrared \citep[e.g.,][]{casasayas-barris_na_2018, casasayas-barris_atmospheric_2019, hoeijmakers_spectral_2019, hoeijmakers_high-resolution_2020, ben-yami_neutral_2020, nugroho_detection_2020, nugroho_searching_2020, stangret_detection_2020, yan_temperature_2020, cabot_detection_2020, cabot_toi-1518b_2021, cont_detection_2021, deibert_detection_2021, kesseli_atomic_2022, merritt_inventory_2021, prinoth_titanium_2022}. On the other hand, the inventory of molecular species in UHJs is lacking. Starting with MASCARA-1 b, CRIRES+ is well suited to fill this gap.
 
 Furthermore, highly irradiated gas giants are predicted to host temperature inversion layers due to the strong optical and UV absorption in the upper atmosphere by TiO, VO \citep{hubeny_possible_2003, fortney_unified_2008} and other atoms and molecules \citep{molliere_model_2015,lothringer_extremely_2018, gandhi_new_2019}. Although these inversion agents have not been extensively detected, temperature inversions have been detected in multiple UHJs, e.g. WASP-33 b \citep{haynes_spectroscopic_2015, nugroho_high-resolution_2017, nugroho_detection_2020}, WASP-121 b \citep{evans_detection_2016}, and WASP-18 b \citep{sheppard_evidence_2017, arcangeli_h_2018}. MASCARA-1 b is a good candidate for an inversion given that the measured temperature using the Spitzer 4.5 $\mu$m-band, probing strong CO features, is significantly higher than the equilibrium temperature \citep{bell_comprehensive_2021}.

In what follows, we outline the science verification observations in section \ref{sec:obs}. We introduce the data calibration and reduction plan in section \ref{sec:reduction}, first giving a detailed overview of both the CR2RES  pipeline and our custom-built ExoRES pipeline and then describing how we use them. Finally, we compare both pipelines and validate the data reduction. We then assess the performance of CRIRES+ in section \ref{sec:performance}, discussing the spectral resolution and quality metrics. We continue with the case study of MASCARA-1 b in section \ref{sec:SV}. Finally, in section \ref{sec:discussion}, we present the summary and discussion. 

\begin{figure}
	\includegraphics[width=\columnwidth]{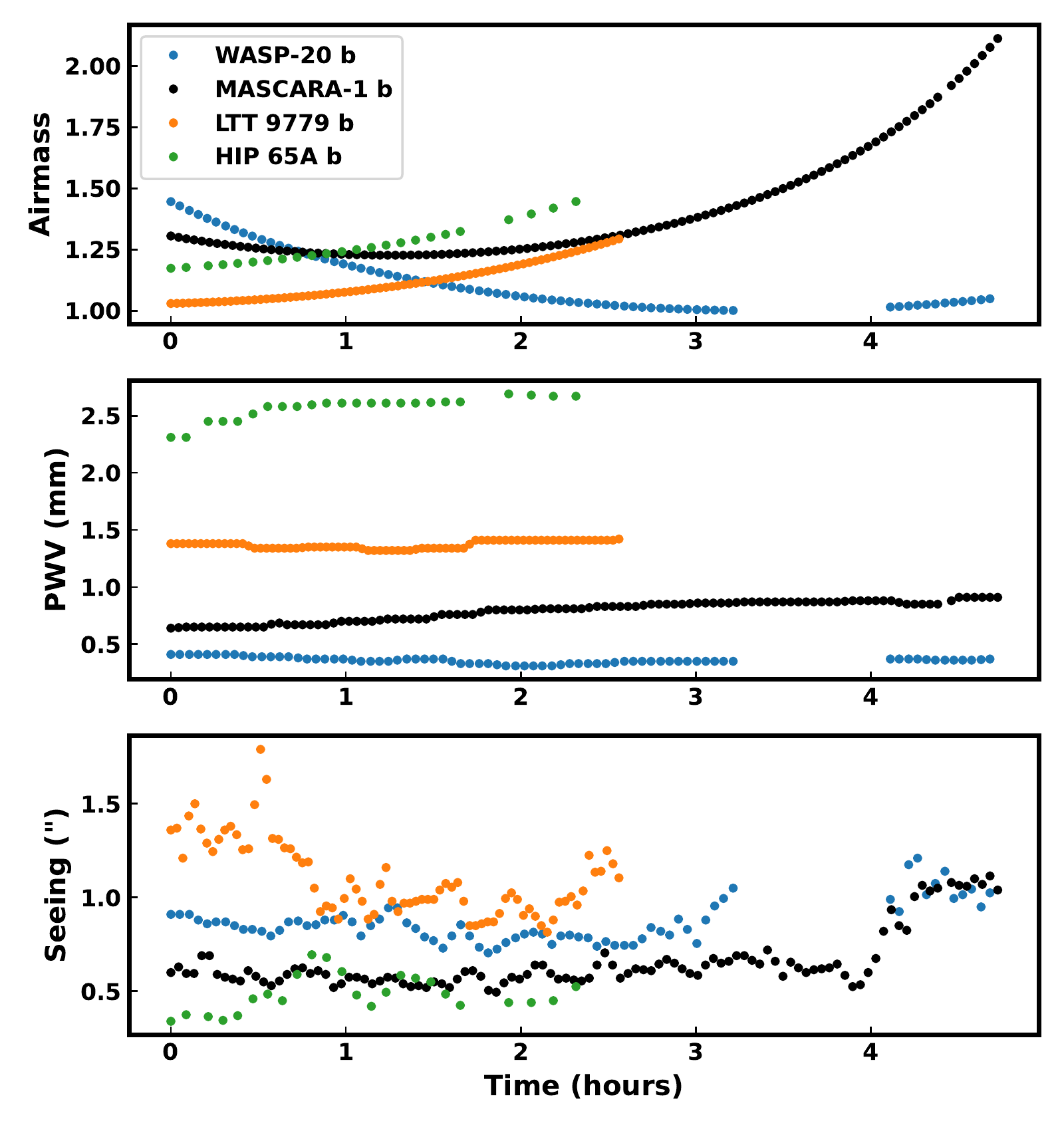}
    \caption{Observing conditions for each target. The time is measured from the start of each observation. The seeing is the average optical seeing at zenith throughout each exposure, as reported by the Paranal Differential Image Motion Monitor (DIMM).}
    \label{fig:conditions}
\end{figure}

\section{Observations} \label{sec:obs}

We focus on observations of transiting exoplanet atmospheres conducted as part of the science verification of CRIRES+ on 15-18 September 2021. Four such targets were observed: [1] Hot Saturn WASP-20 b (Program ID: 107.22SX.001, PI: Maimone), [2] Ultra Hot Jupiter MASCARA-1 b (Program ID: 107.22TQ.001, PI: Gibson), [3] Hot Jupiter HIP 65A b (Program ID: 107.22U7.001, PI: Nagel), and [4] Hot Neptune LTT 9779 b (Program ID: 107.22TE.001, PI: Ramirez Reyes). The observations are publicly available from the ESO Science Archive. The observation of MASCARA-1 b targeted the thermal emission spectrum of the planet dayside, while the three other observations were conducted during transit. The HIP 65A b dataset is specifically aimed at evaluating the radial velocity performance of CRIRES+ while the remaining three datasets are to serve as atmospheric characterisation benchmarks. Our goal is to assess the quality of these observations. All observations were made in the K-band, but with varying wavelength settings and slit sizes. Three of the observations employed the Multi Applications Curvature Adaptive Optics (MACAO), offering near-diffraction-limited performance. The exception is the observation of HIP 65A b, which did not employ the adaptive optics in order to attain a more uniform slit illumination. Moreover, the short gas cell was used for the observation of HIP 65A b to provide a stable wavelength reference. An overview of the observations is shown in table \ref{tab:obs}.

The observing conditions can be seen in figure \ref{fig:conditions}. The airmass remains low throughout all observations, except towards the end of the MASCARA-1 b observation, which caused the S/N to drop slightly (see left panel of figure \ref{fig:SNR_spectra}). For all but the observation of HIP 65A b, the precipitable water vapour (PWV) stays well below the median PWV of 2.5 mm at Paranal. The recommended PWV constraint in the K-band is 2-4 mm, which all observations fulfil. The seeing was excellent during the MASCARA-1 b and HIP 65A b observations, while average to bad during the observations of WASP-20 b and LTT 9779 b.

Depending on the wavelength setting in the K-band, 6-7 orders span the three detectors of CRIRES+. The dispersion in the K-band corresponds to around $\sim$0.9 km s$^{-1}$ per pixel, while the pixel scale in terms of angular size is 0.059". CRIRES+ has two fixed slits, 0.2" and 0.4" in width, that are expected to give a minimum resolving power of ~80,000 and ~40,000, respectively (CRIRES+ User Manual (P109.4)). During good atmospheric conditions and with adaptive optics, the full-width half maximum (FWHM) of the stellar point spread function (PSF) can be less than the width of the slit. We show how this influences the resolution in section \ref{sec:resolution}.

For each target, we use calibration frames taken close in time to the observation. The exception is the high S/N flat field that is taken every 90 days in each wavelength setting, except the M-band. To make our findings easier to replicate, we highlight the calibration files used for the MASCARA-1 b observation in table \ref{tab:calibration}.

\begin{table}
\caption{Calibration data used for the MASCARA-1 b observation.}
\movetableright=-0.5in
\begin{tabular}{lllll}
\hline
Type & Exp. Time &  N$_\mathrm{EXP}$ & Time of first exposure \\ \hline
Dark    & 3 $\times$ 1.427 s    & 3    & 2021-10-12 11:28:16  \\
Dark    & 3 $\times$ 30 s     & 3    & 2021-09-16 11:57:09  \\
Dark    & 3 $\times$ 30 s    & 3    & 2021-09-17 11:36:03  \\
Dark    & 3 $\times$ 60 s    & 3    & 2021-09-17 11:51:05  \\
Flat    & 50 $\times$ 1.427 s    & 5    & 2021-10-04 19:16:24  \\
FPET    & 3 $\times$ 60 s   & 1    & 2021-09-17 11:31:08  \\ \hline
\end{tabular}
\label{tab:calibration}
\end{table}

\section{Data calibration and reduction} \label{sec:reduction}

Here we outline two data reduction strategies, using the official ESO CR2RES reduction pipeline (version 1.0.5) and our custom-built reduction pipeline \pipeline. The initial calibration procedure for CRIRES+ is similar to old CRIRES, and involve (a) subtraction of the dark current, detector glow, sky background, and eventual optical ghosts; (b) non-linear correction, (c) and the correction of pixel-to-pixel sensitivity variations. In staring mode, the first step is achieved by first subtracting an average dark frame with matching exposure time (DIT) and wavelength setting\footnote{This is needed as the thermal background from the instrument itself changes depending on wavelength setting.} from each science frame. For nodding observations, the sky background and ghosts are removed, in addition to the dark current and detector glow, by subtracting two nodding pairs. Finally, the spectra are extracted from the 2D images. See the CRIRES+ Pipeline Manual (0.9.9) for more details\footnote{\href{https://ftp.eso.org/pub/dfs/pipelines/instruments/cr2res/cr2re-pipeline-manual-0.9.9.pdf}{https://ftp.eso.org/pub/dfs/pipelines/instruments/cr2res/cr2re-pipeline-manual-0.9.9.pdf}}.

At the extraction step, the details start to differ from old CRIRES due to the new cross-disperser unit. Instead of one spectrum per detector, there are now up to 10 spectral orders that are spaced unevenly across the three detectors, while not being perfectly horizontal, and where the projection of the slit is tilted (i.e. not vertical on the detectors) and changing with wavelength. The order tracing on the detectors and the wavelength-dependent slit image curvature is estimated using a flat frame and the Fabry-P\'erot Etalon (FPET), respectively. The new FPET is used to create lines that are evenly spaced in frequency by means of interference in the Y- to K-bands. These regularly spaced lines can be used for wavelength calibration (together with an absolute reference) and for tracking the curvature of the slit image. In the L- and M-band, the slit image curvature could potentially be tracked by the sky emission in the science frames. 

When it comes to wavelength calibration, we perform the wavelength calibration uniformly with \pipeline, using the imprinted telluric lines as a wavelength reference. 

\subsection{CR2RES reduction pipeline}

Depending on the user's needs, there are a few ways to reduce the raw data using the CR2RES pipeline. Our strategy is to manually reduce the calibration data. Once processed calibration data has been generated, the recipe used for calibration and extraction is either \texttt{cr2res\_obs\_staring} or \texttt{cr2res\_obs\_nodding}, depending on observation type. 

\subsubsection{Calibration and order tracing}

Here we outline how to process the calibration data step-by-step. First, we use the CR2RES recipe \texttt{cr2res\_cal\_dark} to combine each set of raw dark frames into master dark frames for every exposure time setting. This script also produced a bad pixel map of hot and dead pixels. The flat field frames are dark subtracted and averaged using the \texttt{cr2res\_util\_calib} recipe with \texttt{collapse=MEAN}, which take as input the raw flat-field frames, the appropriate master dark, and the bad pixel map. Using the resulting master flat-field frame, we perform the order tracing using the \texttt{cr2res\_util\_trace} recipe. The order tracing is saved in a so-called TraceWave-table (TW table). The TW table from \texttt{cr2res\_util\_trace} is passed to the \texttt{cr2res\_util\_slit\_curv} recipe, together with the raw FPET frame, producing an updated TW table with information about both the order tracing and slit image curvature. Next, we extract the blaze function and a model of the flat field frame using the \texttt{cr2res\_util\_extract} recipe, where we input the updated TW table in addition to the master flat-field frame. The normalised flat field frame and an additional bad pixel map are created using the \texttt{cr2res\_util\_normflat} recipe, which divides the master flat frame by the extracted flat field model from above. The normalised flat field represents pixel-to-pixel variations and contains only values close to unity, ensuring that we do not magnify the noise in low-signal regions and that the Poisson statistics of the science frames are preserved. 

If the option \texttt{subtract\_nolight\_rows} is to be set to false during extraction (default), the two bad pixel maps can be merged using the \texttt{cr2res\_util\_bpm\_merge} recipe. Setting \texttt{subtract\_nolight\_rows=TRUE} will subtract a vertical median of the bottom 40 rows, which are baffled to not receive light. This corrects for readout artifacts in the form of vertical bands, which are visible in many of the raw frames. However, since low flux pixels in the flat field are identified as bad pixels by default, including the option \texttt{subtract\_nolight\_rows=TRUE} will produce an error when using the bad pixel map from \texttt{cr2res\_util\_normflat}.

At the time of writing, non-linear corrections are not supported given that some of the calibration data have not yet been produced. This should not be a problem as the flux of the spectra we consider is low enough for the pixel shot noise to dominate over the non-linearity error (which should be below 1\%, according to the CRIRES+ User Manual (P109.4)). However, the lack of non-linear correction might worsen the flat field correction since the flat frames are obtained at higher flux. Despite this, we achieve robust molecular detections in the atmosphere of MASCARA-1 b, as shown in section \ref{sec:SV}.

\subsubsection{Spectrum extraction}

For observations in staring mode, the recipe used to calibrate and extract the science spectra is called \texttt{cr2res\_obs\_staring}. We provide the recipe with the master dark and the normalised flat field, a bad pixel map, a TW table, and a science frame. The method is the same for nodding observations, except that a nodding pair is given to the extraction recipe \texttt{cr2res\_obs\_nodding}. Note that both these recipes return an average spectrum if multiple exposures are given, meaning that for time-series data, each observed spectrum (or nodding pair) needs to be run separately. Finally, the blaze function can be corrected for by the \texttt{cr2res\_util\_splice} recipe, which returns the blaze function corrected spectrum on a common wavelength grid. 

\begin{figure}
	\includegraphics[width=\columnwidth]{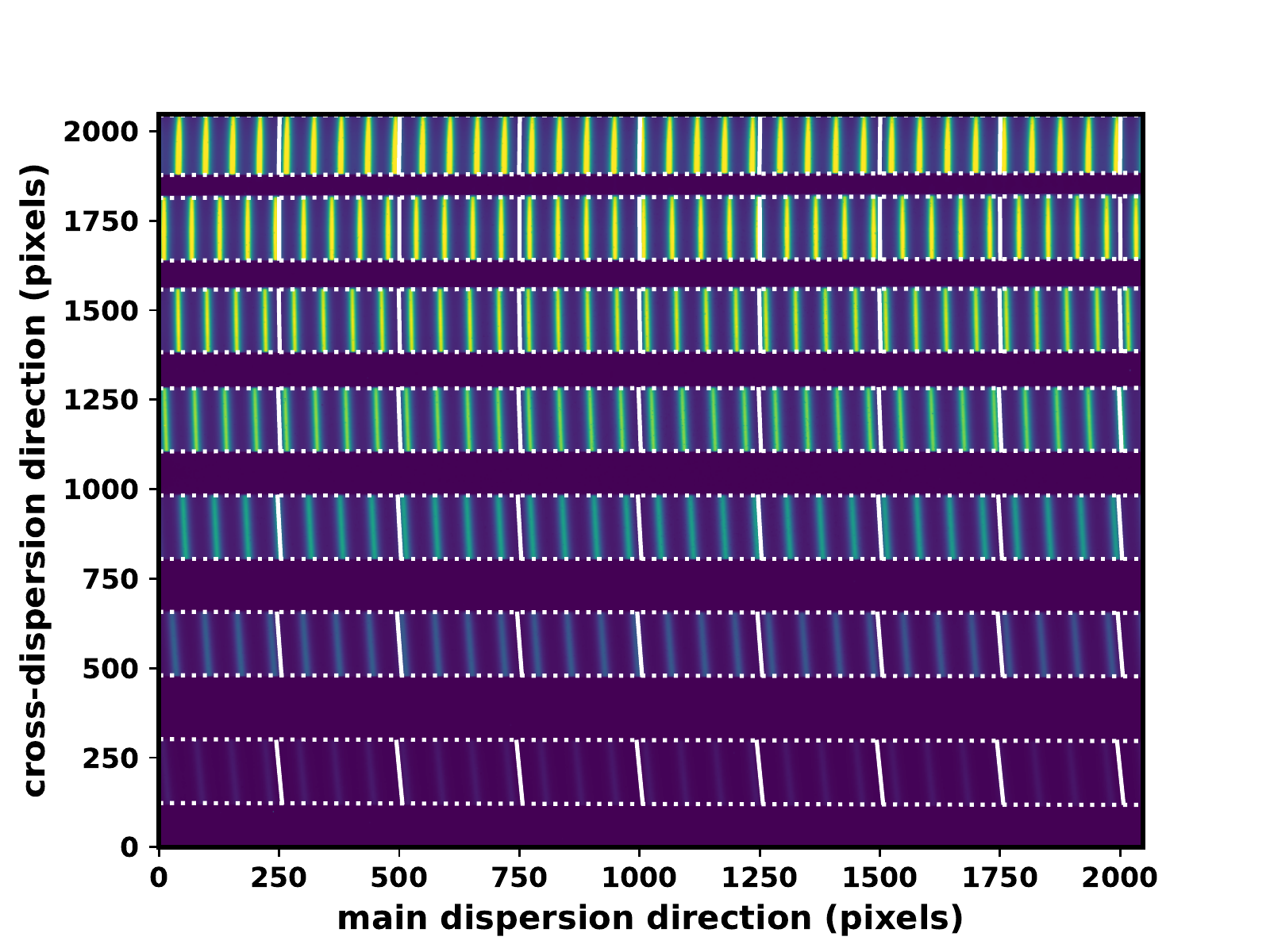}
    \caption{A raw Fabry-P\'erot Etalon frame of detector 1 in the K2166 band, overlaid with horizontal dotted lines representing the order tracing and solid lines representing the tilt of the slit image. For similar figures with other wavelength settings see the CRIRES+ Pipeline Manual (0.9.9).}
    \label{fig:fpet}
\end{figure}

As for the extraction algorithm itself, the basic principle is to decompose the 2D image of each order into an outer product of the slit illumination function (i.e. the profile) and the spectrum, hence making a model of the data, while accounting for optical aberrations such as the tilt of the spectrum and the wavelength dependant slit image curvature. The extraction is done in pieces, by dividing the image into several swaths with widths determined by the \texttt{extract\_swath\_width} parameter (default is 800 pixels), in which the profile and slit illumination function is assumed not to vary. Smaller swath widths allow for more accurate modelling of aberrations but require high S/N spectra. During extraction, the slit illumination function is oversampled to accurately capture the slit image curvature. If the oversampling factor is set too small, the extracted spectra might show artifacts. To mitigate this effect, the oversampling factor can be set by the parameter \texttt{extract\_oversample}, which is 5 by default. The CRIRES+ Pipeline Manual (0.9.9) suggest a factor of around 10-12 in the worst case, hence we choose a value of 12 to be on the safe side. After extraction, the residuals of the observed image and the model is computed and quadratically added to the pixel uncertainties (from shot noise, etc.), which is then converted to spectral bin uncertainties using the extracted profile from each swath. Full details about the extraction algorithm is found in \cite{piskunov_optimal_2021}. 

Given the novelty of CRIRES+, we run the extraction recipe with multiple parameter variations for the sake of robustness.

\subsection{\pipeline reduction pipeline}

We now discuss our custom-built CRIRES+ data reduction pipeline called \pipeline. The underlying extraction algorithm in \pipeline is different from the one used by the CR2RES pipeline, making a comparison particularly interesting.

\subsubsection{Calibration and order tracing}

Again, we start by making the master dark frame. We produce a bad pixel map by applying iterative sigma clipping to exclude 5$\sigma$ outliers, resulting in $\sim 2 \%$ bad pixels across the three detectors. Next, we dark subtract and average the raw flat frames and use the gradient of the resulting master flat, in the cross-dispersion direction, to fit the edges of the illuminated orders using a quadratic polynomial. We exclude orders with a height smaller than 125 pixels. Using this order tracing information, we extract the curvature of the slit image from the FPET frame (which we calibrate before using). Per order and detector, we fit the centres of the regularly spaced lines for each row to the line centres at the middle of each order, effectively mapping the curved grid to a regular grid. The polynomial model of the slit image curvature that we use is
\begin{equation}
    (x',\, y')  = \left(\sum\limits_{i=0}^2 \sum\limits_{j=0}^2 c_{ij}^{(m)}x^i y^j\,, y\right)\,,
\end{equation}
which maps each pixel in order $m$ to a new grid where all emission lines are vertical. The fitting is iterated to remove the influence of badly fitted lines. We obtain residuals smaller than 0.1 pixels. The FPET frame, overlaid with the order tracing and slit image curvature is shown in figure \ref{fig:fpet}.

Next, we extract the blaze function from the flat field and use this to normalise the flat field image. The blaze function is extracted while taking into account the slit image curvature by using the polynomial fit from above to track which pixels correspond to each wavelength bin. In general, such a curved wavelength bin (with a width of one pixel) intersect two pixels for each position in the spatial direction. We assign weights to these intersecting pixels that are proportional to the area of the intersection of the curved wavelength bin and each pixel. Using these weights, we iterative sigma clip any outliers from the weighted average of the pixels intersecting each curved wavelength bin (rejecting 5$\sigma$ outliers). The final average is the value of the blaze function in a particular bin. Using the same weights for every pixel in the flat field, we construct a model of the flat field where we fill each curved wavelength bin with the value of the blaze function at that position. The final normalised flat field is produced by dividing the average flat field image by this blaze function model image. Compared to the CR2RES pipeline, we do not normalise the flat field in the spatial direction (i.e. cross-dispersion direction), as we found the flat field to be reasonably uniform once the blaze function was used for normalisation.

\begin{figure}
	\includegraphics[width=\columnwidth]{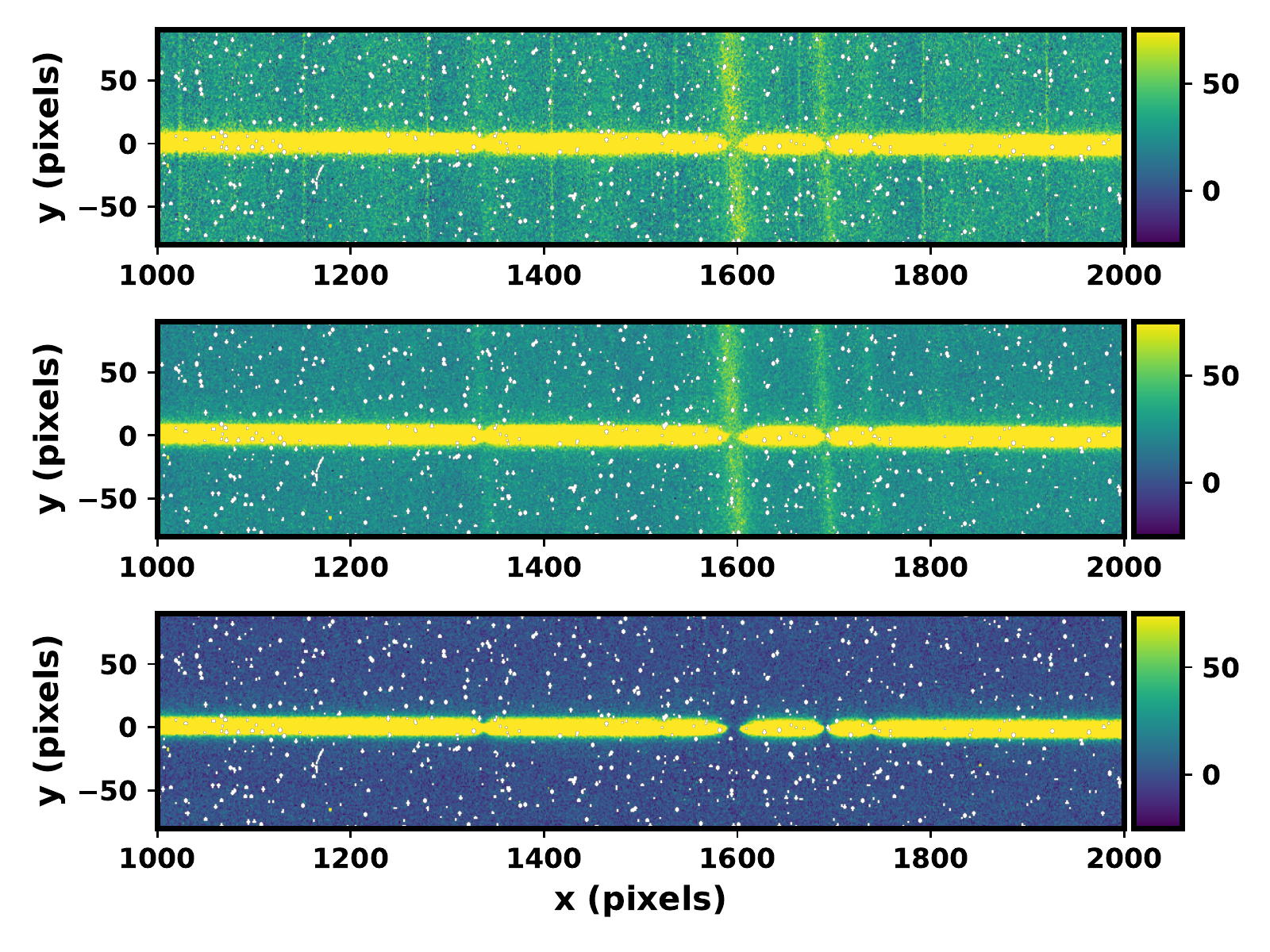}
    \caption{Illustration of the effects of calibration and sky emission subtraction. The top panel shows raw data from the MASCARA-1 b dataset of a section of one order. Bad pixels are marked in white. Sky emission lines appear as tilted lines with excess flux. The vertical bands are detector readout artifacts, some of which remain after dark subtraction. The middle panel shows the same data that has been calibrated and where the vertical bands have been estimated and removed by subtraction (before dividing by the flat field). The bottom panel shows the result of sky emission subtraction using the \pipeline pipeline. The units are ADUs.}
    \label{fig:initial_cal}
\end{figure}

Moving on to the science frames. For staring observations, we begin by dark subtraction. We then estimate and subtract readout artifacts in the form of vertical bands. These bands are estimated by the median of each column, using only the area not illuminated by any order. This approach differs from the method used by CR2RES via \texttt{subtract\_nolight\_rows}\footnote{After submitting the present work we came to know that the option to subtract scattered light and readout artifacts is now included in revised version 1.1.4 of the CR2RES pipeline using the \texttt{subtract\_interorder\_column} setting.}. We find that the latter approach leaves still faint visible bands that can be removed by including more pixels in the estimate of the background. Finally, we complete the calibration by applying the flat field correction and multiplying the data by a factor of NDIT, the number of sub-exposures, such that each pixel count does not correspond to an average. Note that we omit the non-linear correction step as the needed type of calibration data has yet to be produced. In principle, the non-linear correction should be the first calibration applied. For nodding observations, we replace the dark subtraction with the nodding pair subtraction. Here we choose to subtract the linear interpolation of the adjacent frames with complementary nodding positions. To maintain the time resolution of the series, we do no not average over the nodding frames. 

We proceed by estimating and subtracting the sky emission $S_{xy}$ in the case of the staring observations. To estimate the sky emission, we use the same technique we used to estimate the blaze function model image, hence taking the slit image curvature into account, but now we mask the region where the spectrum is located. Figure \ref{fig:initial_cal} shows an example of raw, calibrated, and sky emission subtracted data.

\subsubsection{Spectrum extraction} \label{sec:our_extraction}

We implement an extraction method that is a variation of the optimal extraction algorithms developed by \cite{horne_optimal_1986, marsh_extraction_1989, piskunov_new_2002, zechmeister_flat-relative_2014, petersburg_extreme-precision_2020}. The idea behind the optimal extraction algorithms is to apply appropriate pixel weights to maximise the S/N of the extracted spectrum while preserving spectrophotometric accuracy. These weights depend on the profile $P_{xy}$, i.e. the point spread function in the spatial direction at each wavelength bin, which thus need to be accurately estimated. A variety of methods have been used in the literature to estimate $P_{xy}$. In this work, we use an empirical profile based on the observed spectrum itself, where the profile of low-S/N regions is obtained through interpolation of neighbouring regions. To begin with, we extract the spectrum $s_{x, \mathrm{rect}}$ from $D_{xy} - S_{xy}$ using a rectangular profile, where $D_{xy}$ is the calibrated data of one order. Since the profile is yet unknown, we corrected bad pixels by linear interpolation in the main dispersion direction. This spectrum gives us an estimate of the profile via $(D_{xy} - S_{xy})/s_{x, \mathrm{rect}}$, which we median filter and bin in the main dispersion direction while ignoring regions with a low signal-to-noise ratio (S/N) or too many bad pixels. The median filtering kernel, the bin size, the cutoff S/N, and the cutoff fraction of bad pixels are all free parameters. We have found that a kernel size of 9 pixels in the main dispersion direction is sufficient to remove cosmic rays and that a bin size of 32 pixels and a cutoff S/N of 5 produce robust profile estimates. The one exception is the WASP-20 b dataset, where we had to mask some regions due to low signal (S/N $<$ 3), as we were unable to estimate a profile in these regions. This is at no loss since a planet signal in these regions would be negligible. Wavelength bins that contain more than 50 \% bad pixels (within 20 pixels of the spectrum) are completely masked from extraction. The final profile $P_{xy}$ is obtained by linearly interpolating the binned profile onto the whole pixel grid and by normalising the sum of each column to unity, i.e. $\sum_y P_{xy} = 1$. Before normalisation, we set the profile to zero 20 pixels away from the profile centre. By using an empirical profile we make few assumptions about the profile shape and the order tilt.

\begin{figure}
	\includegraphics[width=\columnwidth]{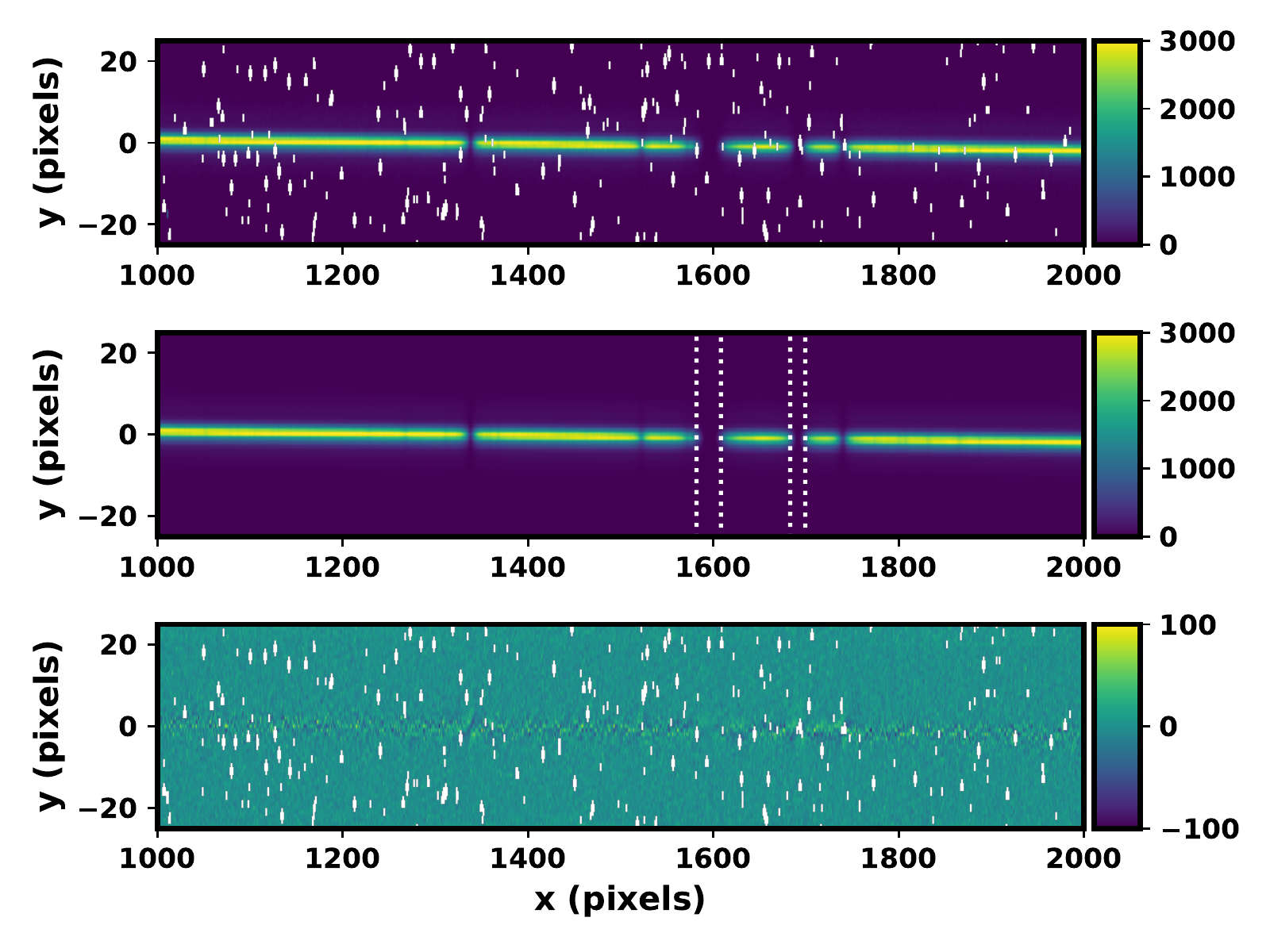}
    \caption{Spectrum extraction. In the top panel is a section of the raw data of an order with bad pixels marked in white. The middle panel is the model of the observation ($S_{xy} + P_{xy} s_x$), where the white dotted lines indicate the regions where the profile is being interpolated due to low S/N. The bottom panel is the residuals between the observation and the model. The units are ADUs.}
    \label{fig:extract_model_spectrum}
\end{figure}

With the profile at hand, the optimally extracted spectrum $s_x$ is obtained by minimising the quantity
\begin{equation} \label{eq:chi2}
    \chi^2 = \sum\limits_{x,y} w_{xy}\left( D_{xy} - S_{xy} - P_{xy} s_x\right)^2\,,
\end{equation}
where the weights $w_{xy} = M_{xy} / \sigma_{xy}^2$ consist of a binary bad pixel mask $M_{xy}$ divided by the pixel variance $\sigma_{xy}^2$ \citep[see][]{zechmeister_flat-relative_2014, petersburg_extreme-precision_2020}. Minimising \eqref{eq:chi2} with respect to $s_x$ yield the optimally extracted spectrum
\begin{equation} \label{eq:s_x}
    s_x = \frac{\sum_y w_{xy} P_{xy} (D_{xy}-S_{xy})}{\sum_y w_{xy} P_{xy}^2}\,,
\end{equation}
which is the same expression derived in \cite{horne_optimal_1986}. Figure \ref{fig:extract_model_spectrum} compares the calibrated data of an observation to the model spectrum constructed using $s_x$. Implicit in this extraction method is that we assume that each wavelength bin is aligned with the pixel grid; however, this is only approximately true due to the slit image curvature. For a typical profile width of a few pixels, we find this approximation sufficient to recover a spectral resolution of $R > 100,000$ in CRIRES+. The extraction algorithm by \cite{piskunov_optimal_2021} does not make this assumption, thus giving the CR2RES pipeline the potential to yield higher resolutions. We discuss this in section \ref{sec:resolution}. 

The noise model we use is
\begin{equation} \label{eq:noise_model}
    \sigma_{xy}^2 = \sigma^2_{xy,\mathrm{bgr}} + \frac{|P_{xy}s_x|}{g}\,,
\end{equation}
where $\sigma^2_{xy,\mathrm{bgr}}$ is the combined variance from readout and sky emission shot noise, and $g$ is the detector gain\footnote{The gain of the three CRIRES+ detectors as provided by the header is 2.15, 2.19, and 2.00 (e$^-$/ADU), respectively.}. Note that we use ADU units and not photoelectrons. For simplicity, we estimate $\sigma^2_{xy,\mathrm{bgr}}$ when we extract the sky emission, by constructing an image corresponding to the weighted variance of the pixels in each curved wavelength bin. We apply the same approach for nodding observations but we mask the negative spectrum.

As proposed by \cite{horne_optimal_1986}, pixels hit with cosmic rays can be iteratively added to $M_{xy}$ by rejecting pixels that do not satisfy $|D_{xy} - S_{xy} - P_{xy} s_x| < \kappa \sigma_{xy}$. Here we choose $\kappa = 5$. For each iteration, we recalculate $s_x$ and $\sigma_{xy}$, but we do not recalculate $S_{xy}$ and $P_{xy}$ since we make sure these are robustly estimated from the start. It is important to only remove one pixel at a time per column since a cosmic ray will increase the flux estimate of the spectrum, hence increasing the value of more than only the affected pixels in the model $S_{xy} + P_{xy} s_x$. Removing only the largest outlier each iteration above $5\sigma$ per column mitigates this effect. We further note that since \eqref{eq:noise_model} depends on the extracted spectrum, we iterate until convergence, which usually only takes a few iterations.

Via error propagation of \eqref{eq:s_x}, we obtain the extraction uncertainties as
\begin{equation} \label{eq:extract_var}
    \mathrm{var}(s_x) = \frac{\chi^2}{N-\nu} \sum_y \left(\frac{\partial s_x}{\partial D_{xy}} \right)^2 \sigma_{xy}^2 = \frac{\chi^2}{N-\nu} \frac{1}{\sum_y w_{xy} P_{xy}^2}\,,
\end{equation}
where we have re-scaled the uncertainties by the reduced $\chi^2$, just like \cite{zechmeister_flat-relative_2014}, where $N = \sum_{x,y} M_{xy}$ is the number of unmasked pixels and $\nu$ is the number of extracted wavelength bins in $s_x$. We note that the reduced $\chi^2$ is typically close to one (calculated where the profile is non-zero), indicating a good model fit. In the case of the MASCARA-1 b dataset, the average reduced $\chi^2$ of all spectra is $1.05 \pm 0.04$. Further note that we have assumed the sky emission $S_{xy}$ and the profile $P_{xy}$ estimates to be noiseless, due to the large number of pixels being used to derive them. 

Finally, we extract the blaze function $f_x$ using the same profile we used to extract the spectrum, i.e. as $f_x = \sum_y P_{xy} F_{xy}$, where $F_{xy}$ is the blaze function model. Since we plan to further normalise the spectra, the normalisation of $f_x$ is at this point unimportant. The blaze corrected spectra and uncertainties are now given by $s_x/f_x$ and $\sqrt{\mathrm{var}(s_x)} / f_x$, respectively.

\subsubsection{Initial wavelength solution}

Our pipeline provides an initial wavelength solution based on the wavelength ranges in the header and the spacing between the lines of the FPET spectra. We extract the FPET spectra at the position of the observed stellar spectrum and at the mid-line of each order. Using the FPET spectra extracted at the mid-line, we construct a (quadratic) polynomial that maps the pixel positions of each line to a wavelength solution where the line spacing is uniform\footnote{We note that in principle, the FPET lines are not evenly spaced in wavelength but rather in frequency. However, within the detector segments of each order, this approximation introduces an error of less than a pixel, which is good enough given that we refine the wavelength solution later.} and where the endpoints correspond to the wavelength ranges in the header. Once we know the wavelengths of each of the FPET lines, we map these to the spectrum along the stellar spectrum. This means that spectra imaged at different positions on the detectors, as during nodding observations, will have a homogeneous wavelength solution, up to systematic wavelength shifts caused by grating drifts and guiding errors.

\subsection{Telluric wavelength calibration}

\begin{figure}
	\includegraphics[width=\columnwidth]{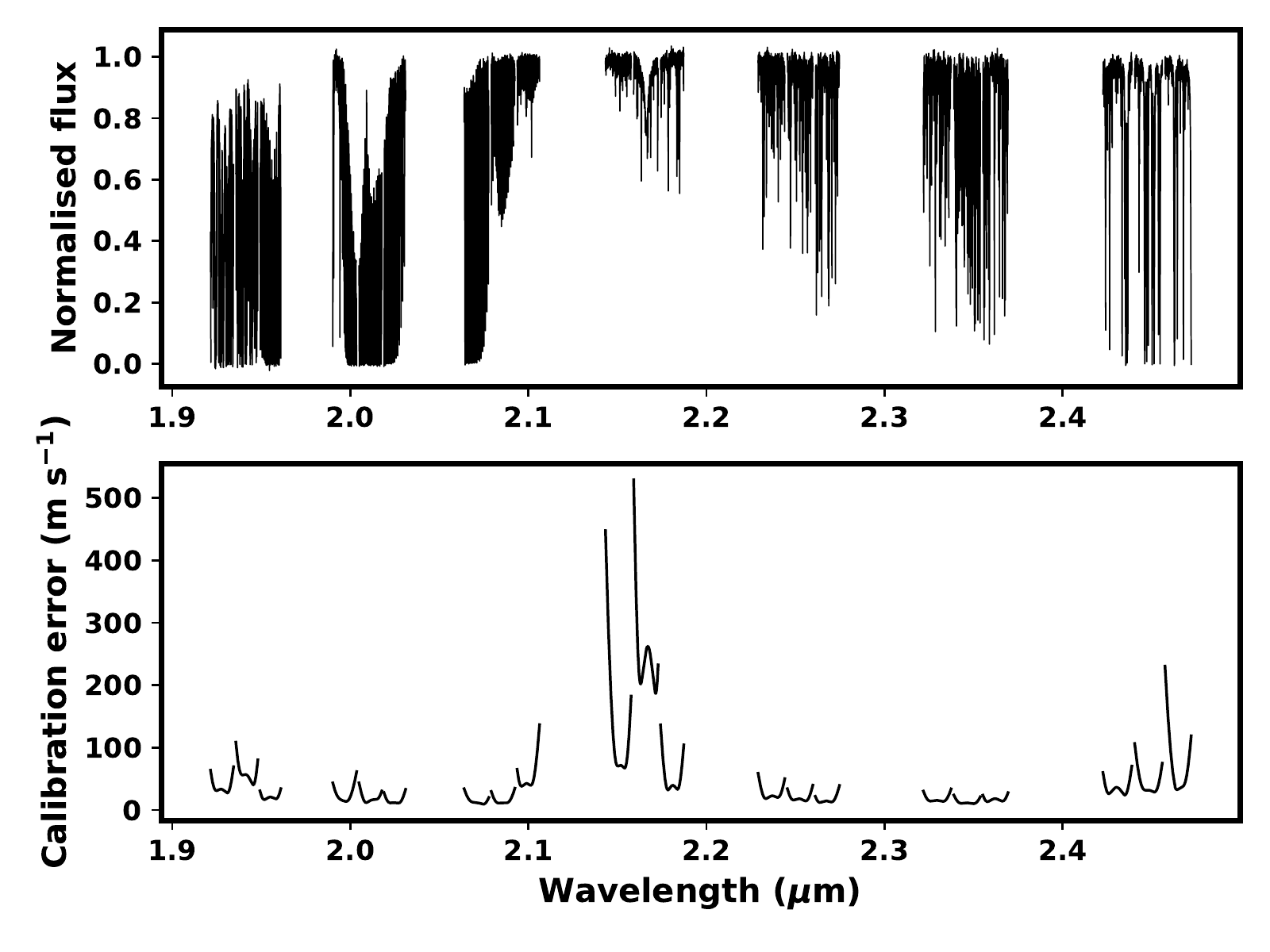}
    \caption{A CRIRES+ K-band spectrum of MASCARA-1 and the resulting wavelength calibration error. The top panel shows an extracted blaze corrected spectrum that has been normalised to match the continuum of a telluric model. Here we used the ExoRES pipeline. The bottom panel shows the wavelength calibration error of the same spectrum as estimated using an MCMC. The data shown here is from the lowest S/N spectrum in the MASCARA-1 b dataset (S/N$\sim$100), thus illustrating the worst-case for this dataset. Regions containing fewer telluric lines produce larger calibration errors. We exclude the two spectra between 2.14 - 2.17 $\mu$m from further analysis due to large wavelength calibration errors.}
    \label{fig:wl_calib}
\end{figure}

We refine the wavelength solution by utilising the imprinted telluric absorption lines, which provide a stable long-term wavelength reference down to 10 m s$^{-1}$ \citep{figueira_evaluating_2010}. We do this by maximising the cross-correlation of the observed spectra with a telluric model \citep{brogi_rotation_2016, chiavassa_planet_2019, webb_weak_2020}. We adopt the telluric model from the ESO Sky Model Calculator \citep{noll_atmospheric_2012}, providing us with a very high resolution ($R\sim10^6$) model. We sample a grid of telluric models in airmass and precipitable water vapour (PWV), which at evaluation, are bilinearly interpolated in log-space. For each detector and order, we assume a quadratic pixel-wavelength solution described by a triplet of evenly spaced points $(\lambda_1, \lambda_2, \lambda_3)$. We convolve the telluric model, sampled with fixed $\Delta \lambda/\lambda$, with a Gaussian to match the resolution of CRIRES+ and set the airmass and PWV to the data given in the header of each spectrum. The wavelength solution is then determined by maximising the cross-correlation function (CCF) as we vary $(\lambda_1, \lambda_2, \lambda_3)$. Stellar lines are masked in the process. We assess the uncertainty of the wavelength solution using a Markov-Chain Monte Carlo (MCMC) routine implemented with the Python package emcee \citep{foreman-mackey_emcee_2013}, where we use the log-likelihood map by \cite{zucker_cross-correlation_2003}. Finally, we re-sample the spectra onto a common wavelength grid using cubic spline interpolation and re-scale the continuum of each spectrum to match the continuum of the telluric model. The wavelength calibration and the estimated calibration errors of the lowest S/N spectrum in the MASCARA-1 b dataset is illustrated in Figure \ref{fig:wl_calib}. Regions with few telluric lines show larger uncertainties.

Given the stability of CRIRES+, an alternative method is to use the average spectrum for wavelength calibration, assuming that the relative alignment between the spectra has already been accounted for by utilising the FPET. Such a  calibration approach could be especially useful for low S/N spectra. 

\begin{figure*}
	\includegraphics[width=2\columnwidth]{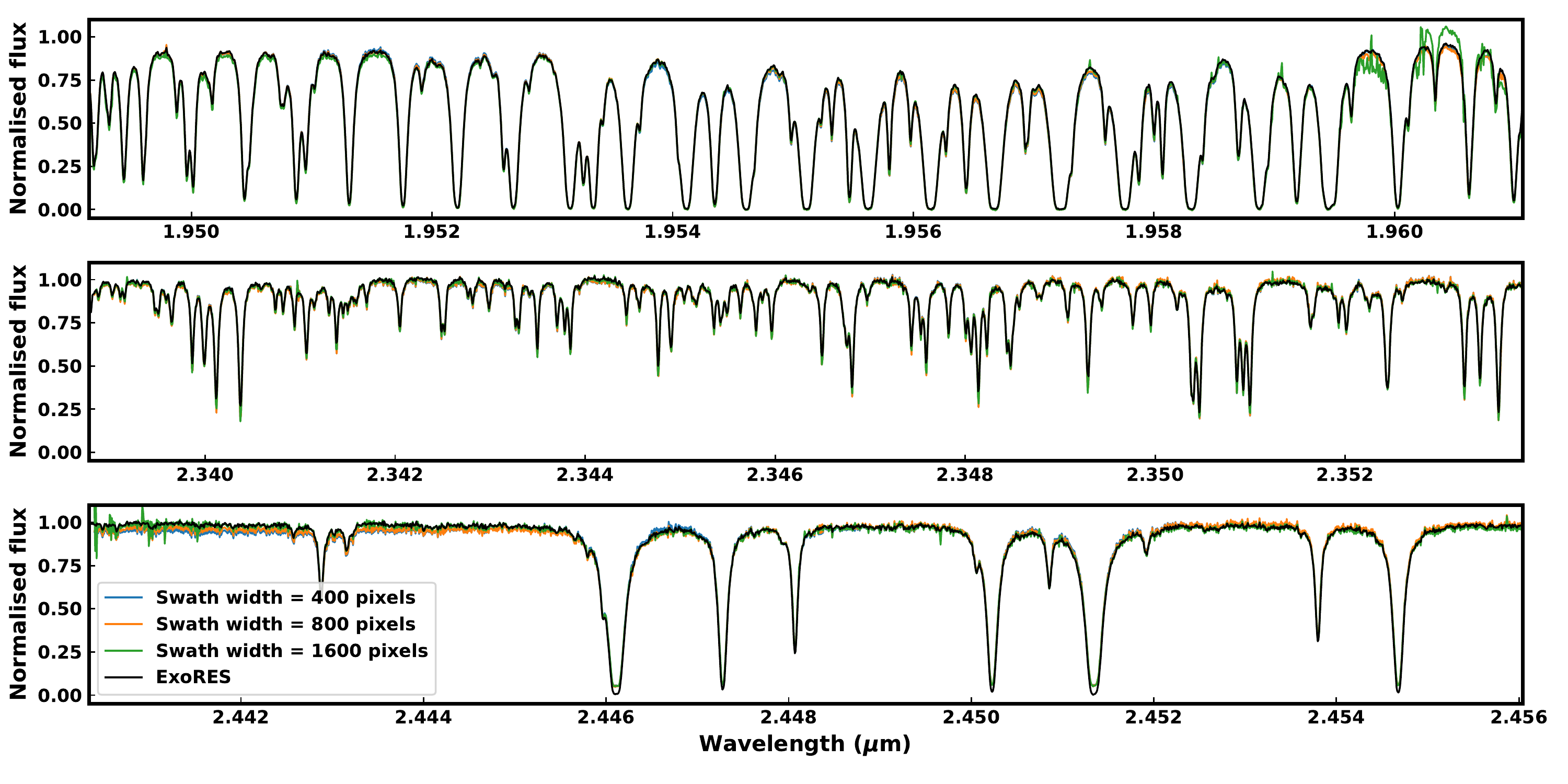}
    \caption{Three regions of a CRIRES+ K-band spectrum of MASCARA-1, showing the general agreement between different extraction methods. The black line is a spectrum extracted using the \pipeline pipeline, while the three other spectra are extracted using the CR2RES pipeline. Here we varied the swath width between 400, 800, and 1600 pixels; shown in blue, orange, and green, respectively. Depending on the swath width, some regions show a varying continuum level and even some spiky artifacts, as can be mainly seen on the right in the top panel. Note that the narrow spectral lines in the middle panel appear slightly deeper in the CR2RES pipeline, while the opposite is true for the saturated lines in the bottom panel.}
    \label{fig:flux}
\end{figure*}

\begin{figure*}
\gridline{\fig{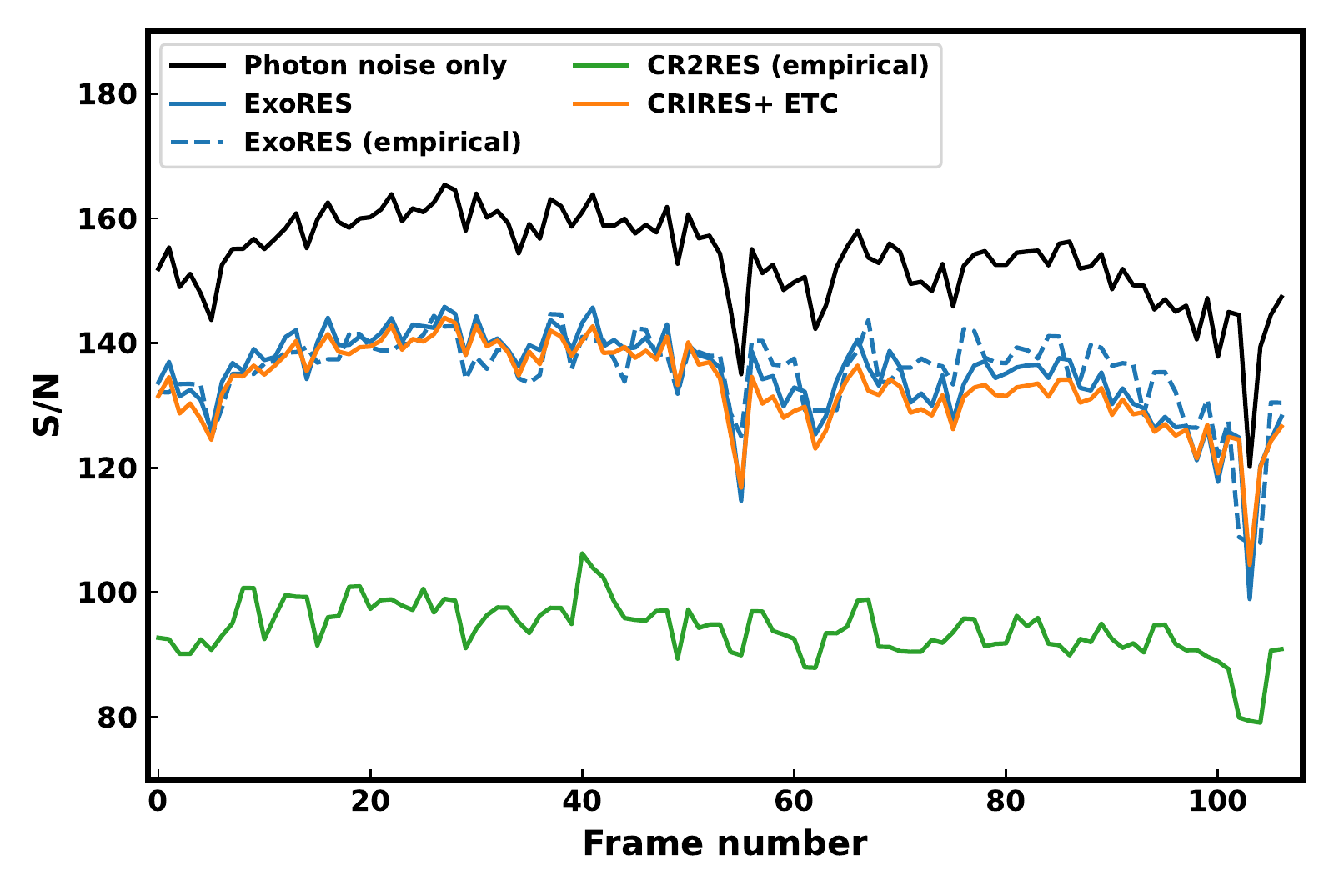}{0.48\textwidth}{}
          \fig{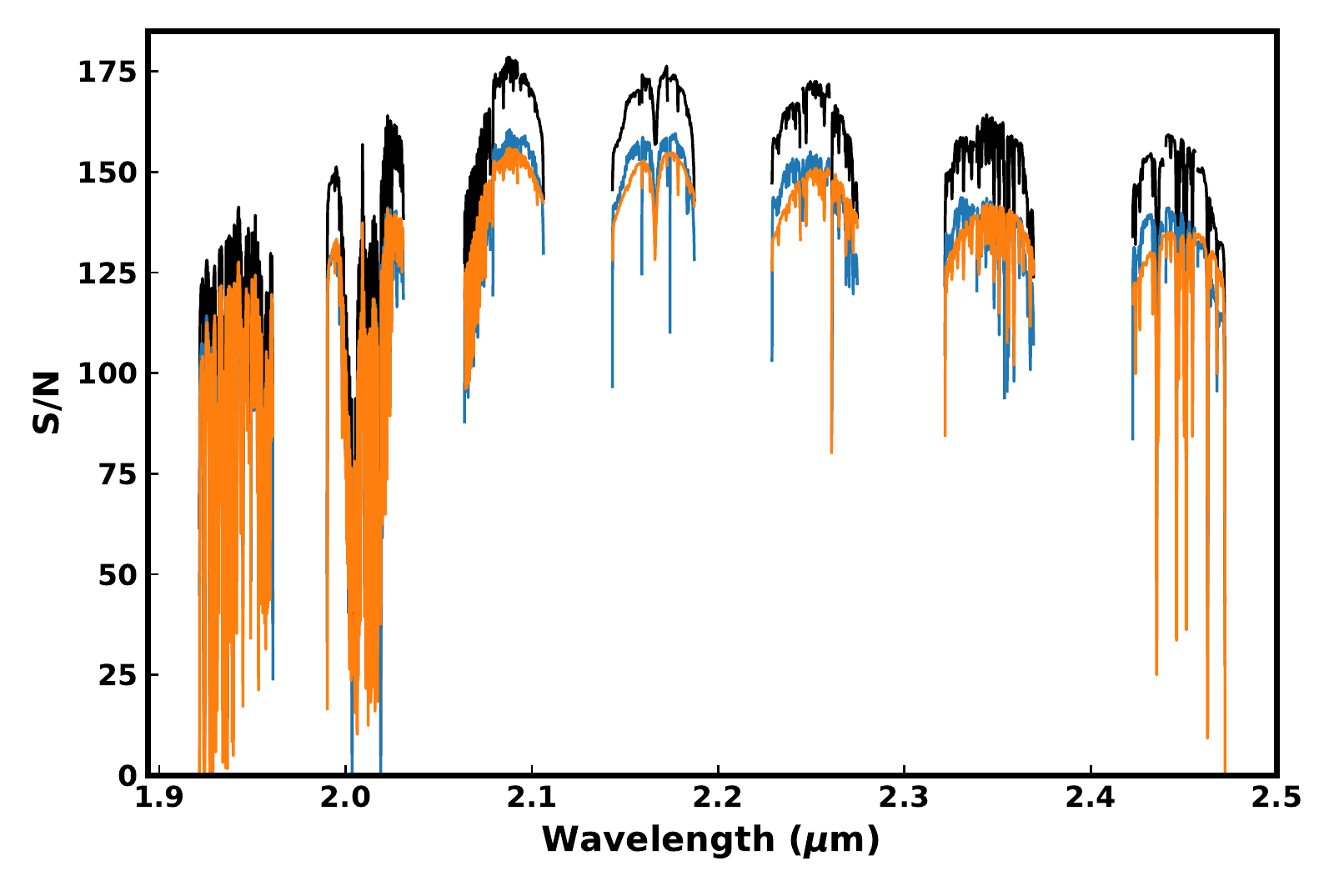}{0.48\textwidth}{}}
\caption{Different S/N estimates of the spectra from the MASCARA-1 b observations as discussed in section \ref{sec:validation}. The left panel shows the median S/N for the different frames as given by both pipelines and the CRIRES+ ETC. The S/N from the ETC is order-wise re-scaled such that the flux given by the ETC matches the observed flux. The right panel shows the average S/N for all orders, with a smoothing  applied for clarity.}
\label{fig:SNR_spectra}
\end{figure*}

\subsection{Validation} \label{sec:validation}
To test the reduction procedures, we analyse the observations of MASCARA-1 b, which has the highest-quality data among the observed targets, using both reduction pipelines. The CR2RES pipeline is run with a range of different extraction parameters to assess their impact. We vary the swath width from 400 to 1600 pixels, utilise the \texttt{subtract\_nolight\_rows} option, and use a smoothing factor along the slit between 1.0 to 3.0.

First, we compare the fluxes of the spectra obtained with the two pipelines. In figure \ref{fig:flux} we show three sections of a blaze-corrected spectrum extracted with the \pipeline pipeline and with the CR2RES pipeline, using a swath width of 400, 800, and 1600 pixels\footnote{Other extraction parameters are set to their default values, apart from the oversampling factor, which is set to 12 throughout this study.}. Overall, the flux is in good agreement, showing only minor deviations. Although these spectra are normalised, we note that the absolute flux agrees as well. In the top panel of figure \ref{fig:flux} we can see some deviations as we increase the swath width.  Taking a detailed look at the spectra, we observe that narrow spectral lines are deeper in the spectra reduced by the CR2RES pipeline as compared to the \pipeline pipeline, reflecting the higher resolution produced by the CR2RES pipeline. We discuss this in section \ref{sec:resolution}. This is most apparent in the middle panel of figure \ref{fig:flux}. At longer wavelengths, i.e. in the lower panel of figure \ref{fig:flux}, we find that our extraction with ExoRES gives deeper saturated telluric lines. This is a result of sky emission subtracting, which is not done by the CR2RES pipeline when extracting staring spectra.

\begin{figure}
	\includegraphics[width=\columnwidth]{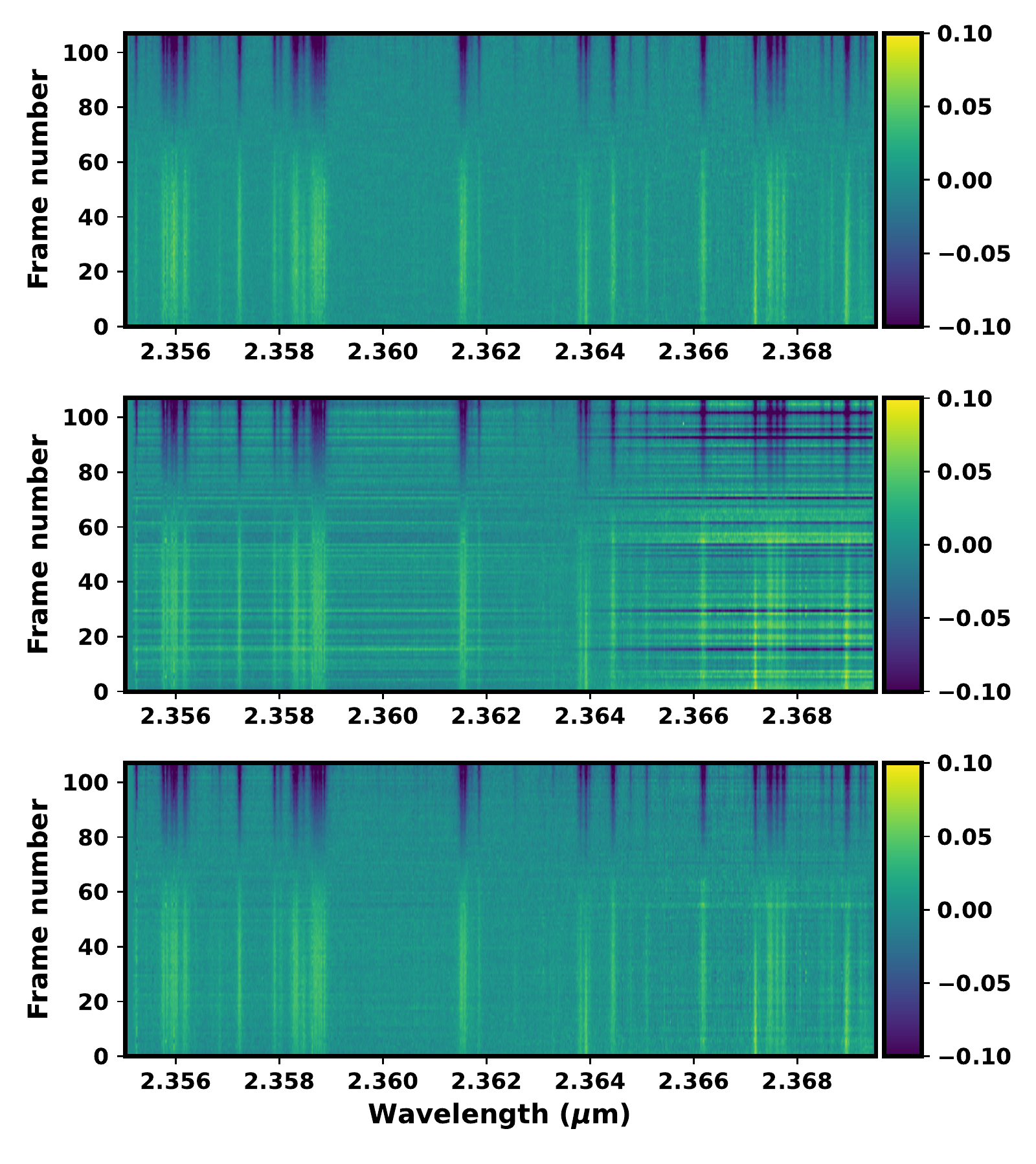}
    \caption{Residuals of a small wavelength interval after subtracting the average spectrum. The top panel shows the residuals obtained from the \pipeline pipeline. The middle panel shows the residuals obtained from the CR2RES pipeline using an 800-pixel swath width and with \texttt{subtract\_nolight\_rows} turned off. The bottom row shows the same as the middle row but with \texttt{subtract\_nolight\_rows} turned on. The data is from the MASCARA-1 b observation. The units are normalised flux, where the continuum level is set to match the telluric transmission before the average spectrum subtracting.}
    \label{fig:flux_time_series}
\end{figure}

Next, we investigate to what extent subtracting readout artifacts affect the extracted spectra. In the CR2RES pipeline, this is controlled by the option \texttt{subtract\_nolight\_rows}. We find that using the \texttt{subtract\_nolight\_rows} option in the CR2RES pipeline improved the continuum homogeneity between spectra in the time-series. We illustrate this in figure \ref{fig:flux_time_series}. Comparing the middle and bottom panels in figure \ref{fig:flux_time_series}, we clearly see excess residuals due to continuum variations when \texttt{subtract\_nolight\_rows} was not used. 

As for the smoothing factor along the slit, we found little to no difference between using a smoothing parameter of 1.0, 2.0, and 3.0. However, we note that other datasets might depend on the extraction parameters differently, e.g. due to different S/N. These results are thus to be seen as a first attempt at characterising the dependence on the extraction parameters in the CR2RES pipeline.

Finally, we assess the S/N per pixel of the derived spectra. In order to conduct an unbiased assessment of the S/N independent of the uncertainty estimates from the pipelines, we first implement an empirical S/N measure which depends only on the flux spectrum. We use a modified version of the robust S/N measure (DER\_SNR) proposed by \cite{Stoehr2008}, which estimates the signal as the median flux and the noise as the third-order order median absolute deviation (MAD) of the flux across the spectral range. The DER\_SNR metric requires the presence of a continuum, which may not be strictly satisfied in the presence of high-density telluric lines. To address this, we use the difference between pairs of spectra (after normalisation) taken close in time, effectively removing most spectral features, and then estimate the MAD on the residuals. To account for the pair-wise subtraction we multiply the resulting S/N by a factor of $\sqrt{2}$, giving the final empirical S/N measure. 

We find that the empirically measured S/N discussed above is in good agreement with the median S/N obtained from the ExoRES pipeline, as shown in figure \ref{fig:SNR_spectra} for MASCARA-1 b. This is also the case for the other targets in our study. We implemented the same comparison for the spectra obtained using the CR2RES pipeline and find that the empirical S/N is higher than the median S/N obtained from the pipeline\footnote{Note that the median S/N we obtain using the CR2RES pipeline agrees with the median S/N found in the header after extraction. The S/N is given by the flux divided by the uncertainty (without any conversion factor), as per the CRIRES+ Pipeline Manual (0.9.9).}. Given this finding and the fact that the flux spectra are comparable between the two pipelines, the empirical S/N is a more reliable measure for comparing the spectra. We find that the empirical S/N achieved by the ExoRES pipeline is higher than that of the CR2RES pipeline, as shown in figure \ref{fig:SNR_spectra}. We found this to be the case for all the extraction parameters we considered for the CR2RES pipeline. It is possible that the S/N from CR2RES could be improved by a better choice of extraction parameters which could be investigated in future studies.

\begin{table*}
\caption{Quality metrics for the observed targets. Here we used the average value for the airmass and seeing.  We report both the seeing from DIMM at Paranal (0.5 $\mu$m, zenith) and the average FWHM of the stellar PSF measured directly on the CRIRES+ detectors. The (median) S/N and resolution ranges are given by the minimum and maximum values. All values are from the \pipeline pipeline.}
\movetableright=-0.8in
\scriptsize
\begin{tabular}{lccccccccc}
\hline
Target   & K magnitude & Spectral type & Exp. Time & Airmass & Seeing  (PSF) & S/N & S/N$_{\mathrm{avg}}$ & $R$ & $R_{\mathrm{avg}}$ \\ \hline
WASP-20 b & 9.4 & F9V & 1 $\times$ 180 s & 1.12 & 0.87" (0.19") & 18 - 33 & 28 & 100,000 - 121,000 & 112,000 \\
MASCARA-1 b & 7.7 & A8 & 5 $\times$ 30 s & 1.41 & 0.65" (0.13") & 99 - 146 & 135 & 106,000 - 118,000 & 114,000 \\
LTT 9779 b & 8.0 & G8 & 1 $\times$ 120 s & 1.12 & 1.10" (0.19") & 45 - 74 & 66 & 91,000 - 114,000 & 105,000 \\ 
HIP 65A b & 8.3 & K4V & 1 $\times$ 300 s & 1.27 & 0.48" (0.34") & 18 - 68 &  44$^{(\mathrm{a})}$ & - & 108,000$^{(\mathrm{b})}$ \\\hline
\end{tabular}
\label{tab:metrics}
\newline
\footnotesize{\textbf{Notes}. $^{(\mathrm{a})}$Here we report the S/N of the exposures using the short gas cell. \\ $^{(\mathrm{b})}$Measured using the last four exposures taken without the gas cell. The telluric fit was inaccurate for the spectra that employed the gas cell. \hfill}
\end{table*}

We also compare the S/N obtained from the different metrics above to that from the CRIRES+ Exposure Time Calculator (ETC). To accurately reproduce the observation with the ETC, we use the average airmass, PWV, and seeing as input; and perform an order-wise re-scaling of the ETC S/N by the square root of the ratio of the observed flux and the flux given by the ETC (i.e. the integrated target signal). This ensures that any difference in S/N between the observed spectra and the ETC is not due to a difference in flux. We find that the re-scaled S/N from the ETC closely matches the S/N from the \pipeline pipeline. Because the S/N from the \pipeline pipeline behaves as expected, agreeing with the empirical measure and the ETC, this is what we report in table \ref{tab:metrics}.

\subsection{Measuring spectral resolution} \label{sec:resolution_method}

In order to assess the on-sky performance of CRIRES+, we measure the resolution by fitting a high-resolution telluric model to the blaze corrected spectra. Our method of determining the resolving power is based on work by \cite{chiavassa_planet_2019}. We again utilise the telluric model from the ESO Sky Model Calculator. The instrumental profile (IP) is approximated as a Gaussian, which we convolve with the telluric model. The airmass and PWV are taken from the header of each spectrum, leading to only one parameter being fitted, the FWHM of the IP. We determine the FWHM of the IP by maximising the cross-correlation between the model and each spectrum. Before fitting, we refine the continuum using the spectrum divided by the telluric model, excluding regions that deviate by more than $5\sigma$. The excluded regions are interpolated and the continuum is smoothed before being used to divide the spectrum. We define the uncertainty of the IP's FWHM as the interval made by $\Delta \ln \mathcal{L} = 0.5$, where we use the cross-correlation to log-likelihood map from \cite{zucker_cross-correlation_2003}.

For robustness, we mask $5\sigma$ outliers to limit the influence of stellar lines and remaining bad pixels. We also perform the fit by minimising the $\chi^2$ using the pixel error from the extraction, which results in a nearly identical fit. Before fitting, we conservatively select orders with the best wavelength calibration and without many saturated broad telluric futures, as these regions contain little information about the IP. In the case of MASCARA-1 b, we only fit the two orders between 2.2-2.4 $\mu$m. The fitting itself is performed individually for each order and detector. We define the final resolution per exposure as the median of these resolution estimates. The typical resolution uncertainty is around 5,000 for the datasets we analysed.

\section{Results: CRIRES+ Performance} \label{sec:performance}

Our primary goal in this study is to assess the performance of CRIRES+ for high-resolution Doppler spectroscopy of exoplanetary atmospheres. Here we assess the performance of CRIRES+ using four targets observed during science verification as discussed above. We focus on two performance metrics: the S/N and the spectral resolution. The observations were conducted using different slit sizes, with and without adaptive optics, and in a variety of atmospheric conditions, thus covering a wide range of observing conditions. CRIRES+ is designed for a resolving a power of $R = 100,000$ when using the 0.2" slit \citep{ramsay_crires_2014}. However, during the commissioning of CRIRES+ the resolving power was measured to be lower than the expected $R = 100,000$. Because of this, the CRIRES+ User Manual (P109.4) states that users should not expect more than $R = 80,000$ for this initial phase of observations. We investigate both the S/N and R achieved for the four targets in our study. The results are shown in table \ref{tab:metrics}.

\subsection{Signal-to-Noise Ratio} \label{sec:SNR}

\begin{figure}
	\includegraphics[width=\columnwidth]{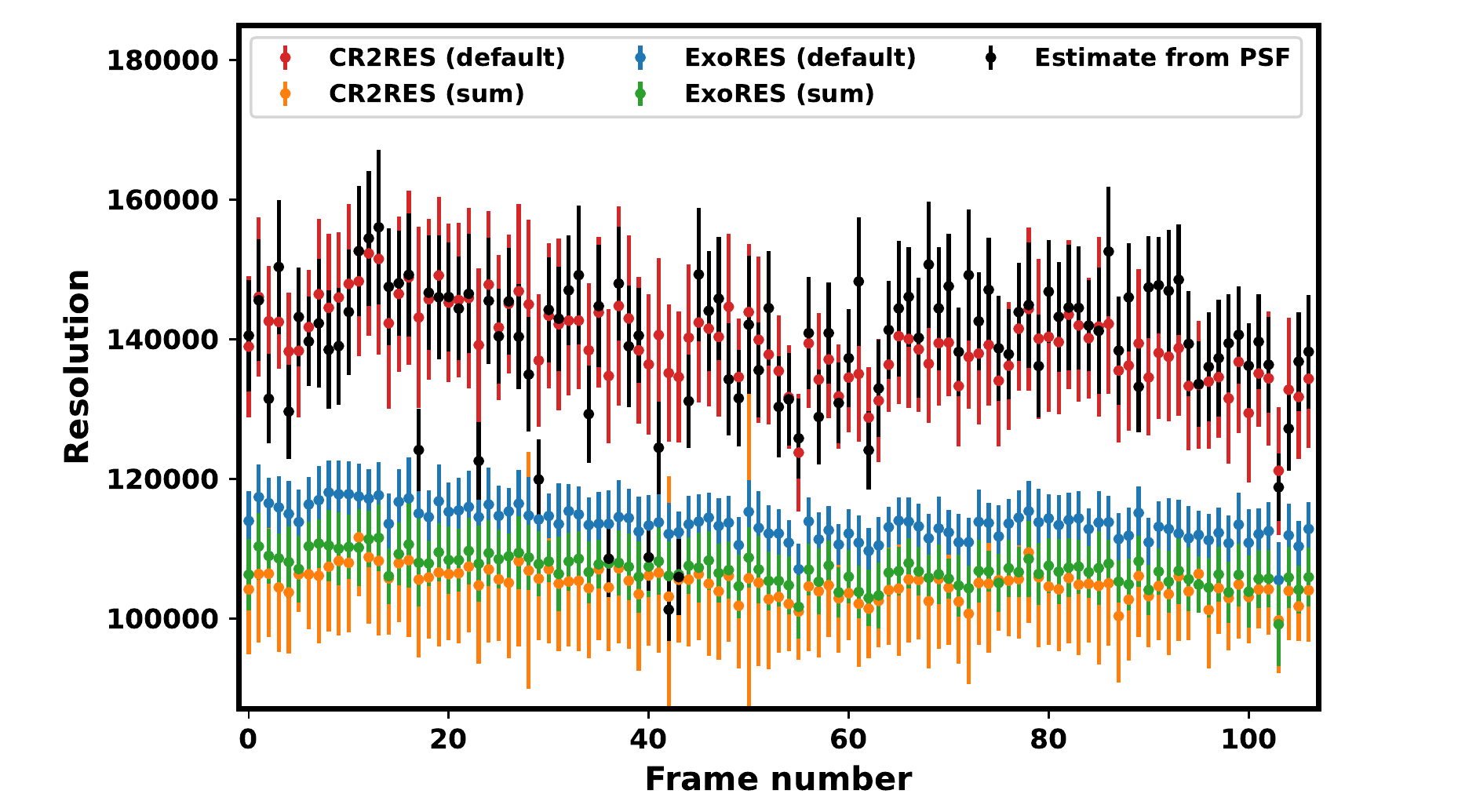}
    \caption{Resolution estimates for the observations of MASCARA-1 b. We compare the resolution estimate obtained by telluric model fitting using different extractions algorithms. In black we show the predicted resolution using the FWHM of the stellar PSF, where each data point is the median from the different orders. The error bars represent the standard deviation across the different orders. We find the resolution to be above $R = 100,000$ in all cases.}
    \label{fig:resolution}
\end{figure}


We first discuss the S/N obtained for the targets across the observations. The S/N is a function of the target brightness, exposure time, and observing conditions. Using our \pipeline pipeline we obtain an average S/N of 28, 135, 66, and 44; for the observations of WASP-20 b, MASCARA-1 b, LTT 9770 b, and HP 65A b, respectively. As expected\footnote{CRIRES+ Pipeline Manual (0.9.9)}, we find that the obtained S/N for these observations is somewhat lower than the theoretical S/N from photon noise alone. The differences in the S/N between the targets are consistent with the corresponding observing conditions. 

Next, we compare the obtained S/N to the S/N predicted by the ETC. We find that the observed S/N (produced by ExoRES) is similar to or higher than the ETC estimate overall. The one exception is the case of HIP 65A b for which the short gas cell was employed, resulting in a lower S/N in the observations compared to ETC estimates. Once we re-scale the ETC output in order to match the observed flux, we find that the S/N from the ETC and the observed spectra are in very good agreement, as can be seen in the case of MASCARA-1 b in figure \ref{fig:SNR_spectra}. 

As discussed in section \ref{sec:validation}, we go on to empirically measure the S/N from the spectra directly using different metrics that are agnostic towards the uncertainty estimate from the pipeline. We find that the S/N obtained with the \pipeline pipeline is in good agreement with the empirically measured S/N from the spectra and with the ETC estimate. Overall, our results indicate that CRIRES+ is performing as expected and providing S/N consistent with the ETC expectations.

\subsection{Spectral resolution} \label{sec:resolution}

Here we discuss the spectral resolution obtained for the targets. When fully illuminating the 0.2" slit, the CRIRES+ spectrograph is designed to deliver a spectral resolution of $R = 100,000$ \citep{ramsay_crires_2014}. However, as mentioned above, commissioning observations with the instrument achieved a somewhat lower resolution of $\sim$80,000. With the present observations, we find that under good atmospheric conditions the spectral resolution of CRIRES+ can reach $R \gtrsim 100,000$ in the K-band. In other words, the spectral resolution of CRIRES+ is not slit-limited for a well-corrected point source.

In the case of MASCARA-1 b, we measure the average resolution to be $R = 114,000$ with spectra reduced by the ExoRES pipeline and $R = 140,000$ using the CR2RES pipeline. This difference is due to the different extraction algorithms used by the pipelines, which we discuss later in this section. We postulate that this very high resolution is a result of the stellar PSF being significantly smaller than the width of the slit, due to the use of adaptive optics. Therefore, we go on to measure the FWHM of the stellar PSF in the spatial direction of each order on the detectors. We find an average FWHM of 0.19", 0.13", 0.19", and 0.34", for the observations of WASP-20 b, MASCARA-1 b, LTT 9779 b, and HIP 65A b, respectively. Since we know the dispersion per pixel we can use the obtained FWHM (for each order) to predict the spectral resolution. This works well for MASCARA-1 b and LTT 9779 b, given that the slit widths are a factor of $\sim2$ larger than the stellar PSF. In figure \ref{fig:resolution}, we show the good agreement between the resolution measured by telluric fitting using spectra from the CR2RES pipeline and the predicted resolution by considering the angular size of the star for MASCARA-1 b\footnote{A few outliers can be seen, which are likely caused by pointing inaccuracies or other optical effects.}. Furthermore, we measure the average resolution of the LTT 9779 b spectra to be $R = 105,000$ as reduced by the ExoRES pipeline and $R = 111,000$ using the CR2RES pipeline. The fact that the resolution is much higher than $R = 40,000$ for LTT 9779 b, which used the wider 0.4" slit, is evidence that the resolution is indeed set by the PSF of the star. 

Given that the spectral resolution of CRIRES+ depends on the size of the stellar PSF, we expect the adaptive optics performance, and in turn, the spectral resolution, to be affected by the seeing during observation. This is demonstrated by comparing the data of MASCARA-1 b, observed during excellent seeing conditions, to that of WASP-20 b and LTT 9779 b, which were observed during worse conditions; resulting in a wider stellar PSF in the latter cases. As expected, this is reflected in the spectral resolution of these observations, resulting in an average resolution of $R = 140,000$, $R = 120,000$ and $R = 111,000$ for for the observations of MASCARA-1 b, WASP-20 b and LTT 9779 b, respectively (when using the CR2RES pipeline). Since a lower resolution is obtained with ExoRES, this trend is not as clear in table~\ref{tab:metrics}.

The relation between the stellar PSF and the spectral resolution is expected to break down when the FWHM of the stellar PSF is similar in size or wider than the slit. To showcase an example where the resolution is slit-limited, we measure the FWHM of the lines from the uranium-neon calibration lamp in the K2166 wavelength setting. Doing this yields a resolution of $R = 82,000 \pm 3,000$ when using the CR2RES pipeline to extract the spectrum. This agrees with the minimum resolving power of $R = 80,000$ as stated by the CRIRES+ User Manual (P109.4). On the other hand, we measure an average resolution of $R = 108,000$ for the observation of HIP 65A b using ExoRES ($R = 109,000$ with the CR2RES pipeline), even though the FWHM of the stellar PSF is on average 0.34", i.e. wider than the 0.2" slit. This may be caused by the non-uniform slit illumination of a point source.

\begin{table}
\caption{Parameters of the MASCARA-1 b system.}
\movetableright=-0.4in
\scriptsize
\begin{tabular}{lc}
\hline
Parameter & Value \\ \hline
Stellar effective temperature & 7490 $\pm$ 150 K $^\mathrm{a}$ \\
Stellar radius & 2.072 $\pm$ 0.022 R$_\odot$ $^\mathrm{a}$ \\
Stellar mass & 1.825 $\pm$ 0.097 M$_\odot$ $^\mathrm{a}$ \\
Planetary equilibrium temperature & 2594.3 $\pm$ 1.6 K $^\mathrm{a}$ \\
Planetary radius & 1.597 $\pm$ 0.019 R$_\mathrm{J}$ $^\mathrm{a}$ \\
Planetary mass & 3.7 $\pm$ 0.9 M$_\mathrm{J}$ $^\mathrm{b}$ \\
Epoch & 2 458 833.488151 $\pm$ 0.00009 BJD $^\mathrm{a}$ \\
Period  &  2.14877381 $\pm$ 0.0000009 days $^\mathrm{a}$  \\
Semi-major axis & 0.040352 $\pm$ 0.000049 au $^\mathrm{a}$  \\
Inclination &  88.45 $\pm$ 0.17 deg $^\mathrm{a}$  \\
Eccentricity &  0.00034 $\pm$ 0.00034 $^\mathrm{a}$  \\ 
RV semi-amplitude ($\Kp$) &  204.2 $\pm$ 0.2 km s$^{-1}$ $^\mathrm{a}$  \\ 
 &  217 $\pm$ 25 km s$^{-1}$ $^\mathrm{b}$  \\
Systemic velocity (V$_{\mathrm{sys}}$) & 8.52 $\pm$ 0.02 or 11.20 $\pm$ 0.08 km s$^{-1}$ $^\mathrm{c}$ \\ \hline
\end{tabular}
\label{tab:parameters}
\newline
\footnotesize{\textbf{Notes}. $^{(\mathrm{a})}$\protect\cite{hooton_spi-ops_2022}. $^{(\mathrm{b})}$\protect\cite{talens_mascara-1_2017}. $^{(\mathrm{c})}$\protect\cite{talens_mascara-1_2017} found a significant offset in the systemic velocity derived from two different datasets.}
\end{table}

We now discuss the difference in spectral resolution produced by the two pipelines, as shown in figure \ref{fig:resolution} for MASCARA-1 b. The resolution we obtain with spectra reduced using the \pipeline pipeline is somewhat lower than that of the CR2RES pipeline. Using the \pipeline pipeline, the average resolution obtained using the MASCARA-1 b dataset is $R = 114,000$. While still higher than $R = 100,000$, this is less than what we find using the CR2RES pipeline. The difference in resolution is due to the different extraction algorithms used by the two pipelines. By not accounting for the slit image curvature during extraction in the ExoRES pipeline, as is being done by the CR2RES pipeline, we lose out on some of the resolution that CRIRES+ can deliver. To verify that this is the case, we utilise the \texttt{cr2res\_util\_extract} recipe in the CR2RES pipeline, which has the option to perform the extraction using different methods. Using the \texttt{SUM} method and an extraction height of 17 pixels, we obtain spectra with an average resolution of $R = 105,000$. Similarly, if we apply the same vertical summation extraction in ExoRES we get an average resolution of $R = 107,000$, illustrating the same effect. The measured resolution using vertical summation extraction is shown in figure \ref{fig:resolution}. This shows that performing the extraction using a vertical sum can degrade the spectral resolution. The difference in resolution between the default \pipeline pipeline and the one obtained by the \texttt{cr2res\_util\_extract} recipe can be attributed to the different extraction heights. The extraction height used in the \pipeline pipeline is effectively the size of the stellar PSF, as discussed in section \ref{sec:our_extraction}, which is only a few pixels. Moreover, we note that the difference in S/N between the ExoRES and CR2RES pipelines, as discussed in section \ref{sec:validation}, might be connected to the fact that the default CR2RES extraction gives a higher spectral resolution.

Overall, we find that across all the targets and the observing conditions considered here the performance of CRIRES+ meets the expected metrics. In particular, under good observing conditions the spectral resolution can reach $R \gtrsim 100,000$ for cases where the PSF is not slit-limited due to accurate corrections from adaptive optics. Similarly, the S/N across the observations are consistent with the ESO CRIRES+ ETC and empirical metrics.

\section{Results: A case study of MASCARA-1 b} \label{sec:SV}

To demonstrate the on-sky performance of CRIRES+ for exoplanet atmospheric characterisation, we conduct a case study of the Ultra Hot Juiter (UHJ) MASCARA-1 b \citep{talens_mascara-1_2017, hooton_spi-ops_2022}. Previous transmission spectroscopy studies in the optical have found no chemical signatures in the planet's atmosphere \citep{stangret_high-resolution_2022}, which was made particularly difficult due to the strong Rossiter-McLaughlin effect when searching for key atomic species. Due to its high temperature, the present CRIRES+ observations of the day-side of MASCARA-1 b in the K-band should make it feasible to detect molecular features in thermal emission. We search for CO and H$_2$O in the planet as these are the prominent oxygen- and carbon-bearing species expected in hot hydrogen-rich atmospheres \citep{madhu2012,moses2013}. The observations of MASCARA-1 b cover orbital phases between 0.32 - 0.42 and consist of 107 spectra with an average S/N of 135. 

\subsection{Outlier removal}

Some bad pixels remain after the spectra have been extracted that can impact the telluric removal and thus worsen the cross-correlation. Since the spectra are normalised and have a common wavelength grid we can detect outliers by performing iterative sigma-clipping on each spectral channel. We reject pixels that deviate more than $5\sigma$ from the mean and replace them with the linear interpolation of its neighbours in time. Furthermore, we do not include the two spectra between 2.14-2.17 $\mu$m as we could not guarantee a sufficiently precise wavelength calibration due to few telluric lines in the region. Figure \ref{fig:wl_calib} show the large calibration error for these spectra. We expect this to have a negligible effect on the cross-correlation analysis as no CO lines exist in this region and only weak H$_2$O lines are present. 

\subsection{Telluric removal} \label{sec:detrending}

To remove the quasi-stationary telluric lines we utilise the principal component analysis (PCA) method first implemented for high-resolution Doppler spectroscopy by \cite{de_kok_detection_2013}. We use singular value decomposition to decompose each flux matrix, i.e. the matrix containing the spectrum time-series from each detector, into a product of singular values and singular vectors. The detrending itself is performed by removing the largest $k$ singular values from each flux matrix, corresponding to subtracting a model that is the best $k$-rank approximation that minimises the sum of the squared residuals. We therefore treat the planetary signal as a small perturbation that should largely survive after removing a low-rank approximation of the data. However, since the above decomposition is purely empirical, there is no guarantee that the planetary signal will not be altered. What remains to do is to robustly select the optimal value of $k$ such that we sufficiently remove the telluric lines and any other systematics, while at the same time making sure that the planet signal remains as intact as possible. Removing too many singular values will degrade the signal. Previous work has shown that selecting detrending parameters, such as the value of $k$, by optimising for the detection significance of an injected signal can produce spurious signals \citep{cabot_robustness_2019, zhang_platon_2020}. Thus, in order to not bias the detection we fix $k$ to be the same for all orders and select the optimal value based on a type of injection test that is insensitive to the injection location. Moreover, we do not mask additional spectral bins, apart from the ones already masked by the reduction. 

\begin{figure}
	\includegraphics[width=\columnwidth]{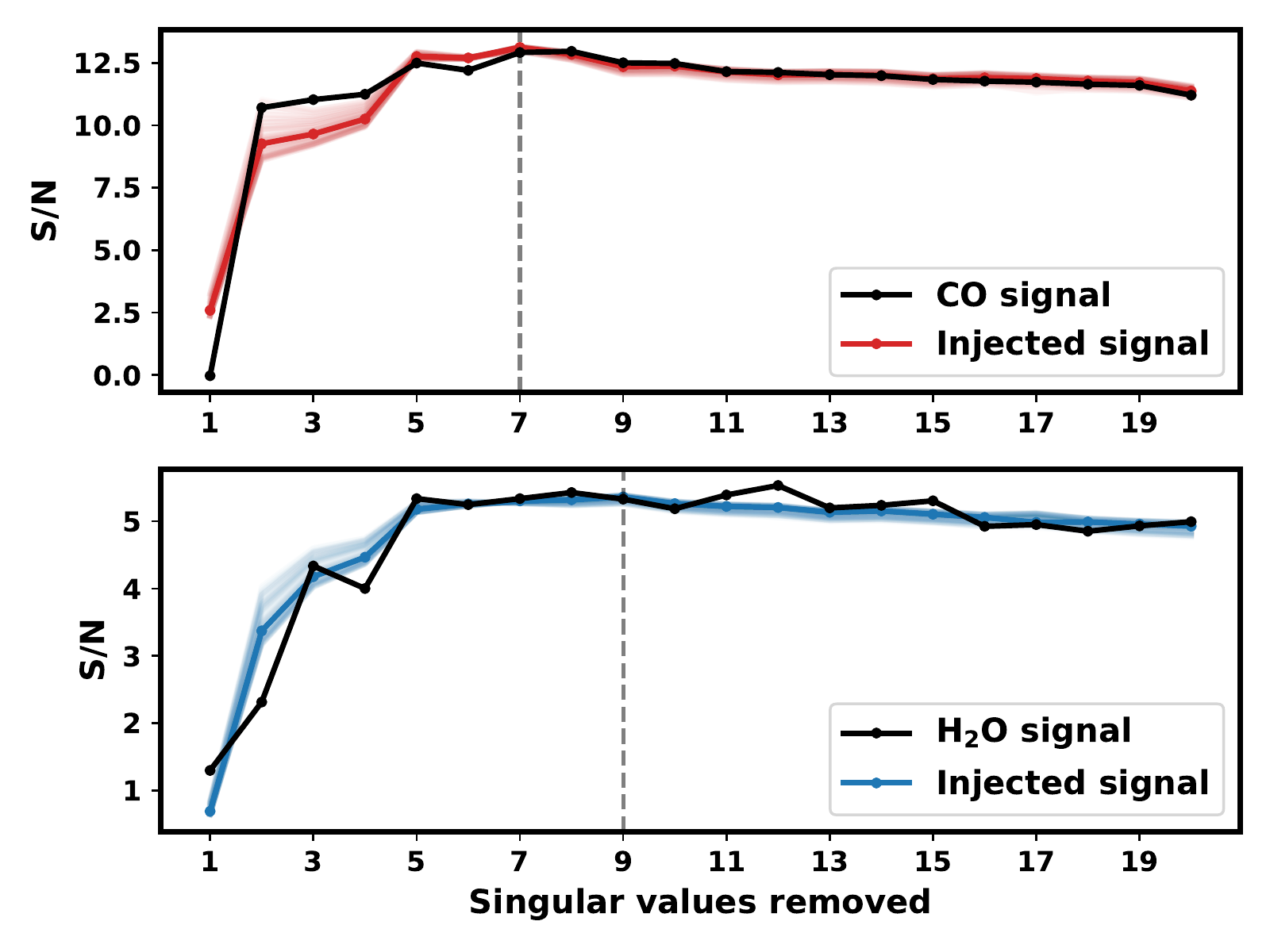}
    \caption{S/N as a function of the number of singular values removed. The blue and red curves show the S/N of CO and H$_2$O model injections at the expected orbital parameters, respectively, while the observed signals are shown in black solid lines. The lighter coloured lines correspond to 100 random injections in the region $\mathrm{K}_{\mathrm{p}} \in [100, 300]$ km s$^{-1}$ and $\mathrm{V}_{\mathrm{sys}} \in [-100, 100]$ km s$^{-1}$. The S/N of the injected models was re-scaled to match the S/N of the real signals. The vertical dashed lines show the optimal number of singular values that maximises the injection S/N at the expected orbital parameters.}
    \label{fig:injection}
\end{figure}

Similar to \cite{hoeijmakers_searching_2018, spring_black_2022}, we select $k$ by maximising
\begin{equation}
    \mathrm{S/N}_{\mathrm{inj}} = \frac{\left(\left.\mathrm{CCF}_{\mathrm{inj+obs}} - \mathrm{CCF}_{\mathrm{obs}} \right) \right|_{\mathrm{K}_{\mathrm{p}}\,, \mathrm{V}_{\mathrm{sys}}} }{\mathrm{std}(\mathrm{CCF}_{\mathrm{obs,masked}})}\,,
\end{equation}
where $\mathrm{CCF}_{\mathrm{inj+obs}}$ is the CCF of the artificially injected signal, where the model has been added to the data, while $\mathrm{CCF}_{\mathrm{obs}}$ is the CCF of the observed data only. The noise is estimated as the standard deviation of $\mathrm{CCF}_{\mathrm{obs}}$ away from the signal, excluding the region $|\Delta V_{\mathrm{sys}}| < 100$ km s$^{-1}$. The final S/N is evaluated at the injection, corresponding to the expected radial velocity semi-amplitude and systemic velocity of the planet, K$_{\mathrm{p}}$ and V$_{\mathrm{sys}}$. Note that our approach to determining $k$ is not very sensitive to the exact K$_{\mathrm{p}}$-V$_{\mathrm{sys}}$ of the injection, as we obtain similar S/N$_{\mathrm{inj}}(k)$ curves when we vary the point of injection. In figure \ref{fig:injection}, we show the S/N analogue of the injected signals as a function of $k$ and compare these to the behaviour of the real signals. The lighter coloured lines correspond to 100 random injections in the region $\mathrm{K}_{\mathrm{p}} \in [100, 300]$ km s$^{-1}$ and $\mathrm{V}_{\mathrm{sys}} \in [-100, 100]$ km s$^{-1}$, illustrating that the optimal number of singular values is not highly dependant on the orbital parameters, at least for close-in planets. Moreover, the fact that the S/N of the CO and H$_2$O model injections show similar behaviour, i.e. increasing sharply for small $k$ and then slowly decreasing at large $k$, further demonstrates the robustness of this method.

Using spectra extracted by the \pipeline pipeline, the optimal $k$ was determined to be 7 and 9 for the CO and H$_2$O models, respectively. We leave it to future studies to investigate the potential benefits and robustness aspects of order-wise optimisation beyond that of \cite{cabot_robustness_2019}. We also adopt SYSREM \citep{tamuz_correcting_2005}, a generalised PCA-based deterending algorithm that is frequently used in the literature \citep[][and more]{birkby_detection_2013, birkby_discovery_2017, nugroho_high-resolution_2017, hawker_evidence_2018, cabot_robustness_2019, sanchez-lopez_water_2019}. The detection significance of both CO and H$_2$O, obtained using SYSREM detrending, is very similar to the significance we get by using the standard PCA approach.

\begin{figure*}
	\includegraphics[width=2\columnwidth]{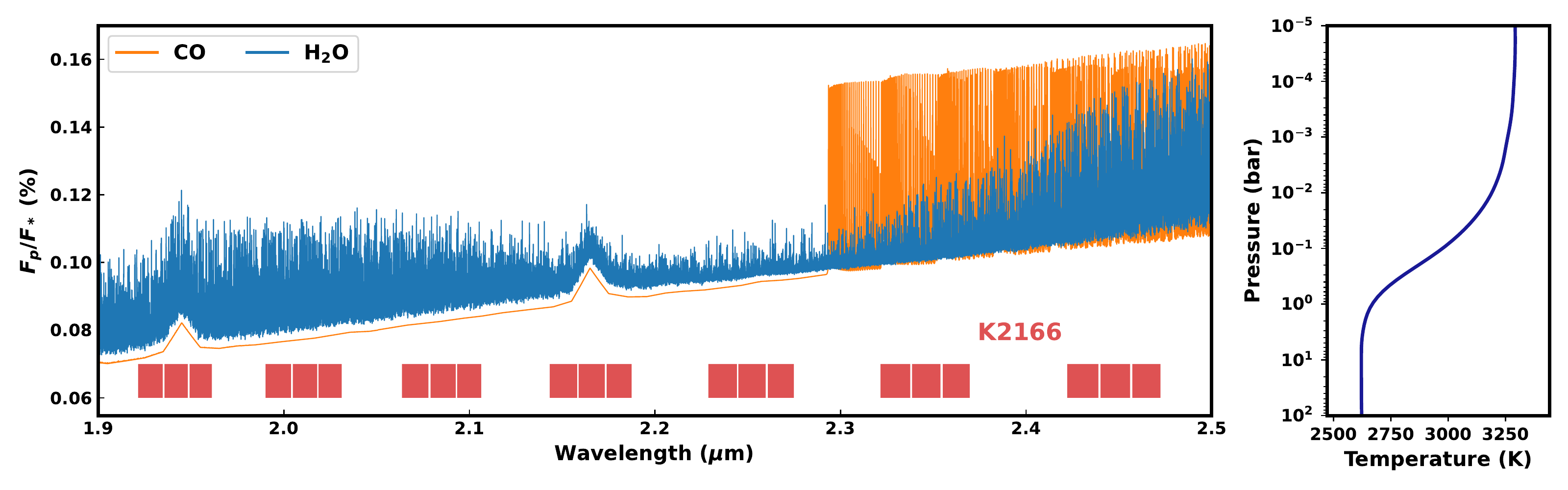}
    \caption{High-resolution model spectra of MASCARA-1 b with contributions from CO and H$_2$O shown individually. The wavelength coverage of the CRIRES+ K-band (setting K2166) used for the observation of MASCARA-1 b is shown in red at the bottom. The right panel shows the pressure-temperature profile used to generate the spectra.}
    \label{fig:models}
\end{figure*}

\subsection{Model atmosphere} \label{sec:model}
In order to cross-correlate with the observed spectra, we compute high-resolution model spectra of MASCARA-1 b using the GENESIS self-consistent atmospheric modelling framework \citep{gandhi_genesis_2017}. The GENESIS code has been used to model irradiated exoplanetary atmospheres across a wide range of conditions \citep{gandhi_genesis_2017,gandhi_new_2019,piette_2020a,piette_2020b}. Here we briefly outline the model set-up and considerations adopted in this work. The model computes the atmospheric temperature structure and radiative transfer self-consistently under the assumption of radiative-convective equilibrium, hydrostatic equilibrium and plane parallel geometry, for a given chemical composition. The atmospheric structure is computed over a pressure range of $10^{-7}-10^3$ bar divided into 100 layers evenly spaced in $\log(P)$. The radiative transfer is solved under the assumption of local thermodynamic equilibrium using the Feautrier method, and the temperature structure is solved iteratively using Rybicki’s method by complete linearization to achieve radiative equilibrium; the formalism of the solvers can be found in \citet{gandhi_genesis_2017} and \citet{hubeny_2014}.

The bulk properties of MASCARA-1 b used for the atmospheric model are shown in table~\ref{tab:parameters}. A Kurucz model is used for the stellar spectrum \citep{Castelli2003}. The GENESIS atmospheric model includes opacity contributions from prominent molecular species expected in hot jupiter atmospheres \citep[e.g.][]{madhu2012,moses2013}, primarily H$_2$O, CO, CH$_4$, CO$_2$, HCN, C$_2$H$_2$, and TiO and collision-induced opacity due to H$_2$-H$_2$ and H$_2$-He. The opacities are adopted from several recent works \citep{gandhi_genesis_2017,gandhi_molecular_2020,piette_2020a}, derived from the HITEMP/HITRAN or ExoMol line list databases for H$_2$O \citep{polyansky2018}, CO \citep{rothman2010,li2015}, CH$_4$ \citep{hargreaves2020}, CO$_2$ \citep{huang2013, huang2017}, HCN \citep{harris2006, barber2014}, C$_2$H$_2$  \citep{Rothman2013,Gordon2017}, TiO \citep{McKemmish2019}. We also include atomic opacity due to Na and K \citep{Burrows2003,gandhi_genesis_2017} and H$_2$-H$_2$ and H$_2$-He CIA \citep{Richard2012}. 

For a canonical model of MASCARA-1 b we assumed the abundances of these chemical species to be those expected for solar elemental composition in thermochemical equilibrium at high temperatures (2500-3000 K) in the planetary photosphere at 10 mbar. The temperature structure is determined by the incident irradiation, the visible and infrared opacity sources, and the day-night energy redistribution, which is expected to be inefficient for highly irradiated hot jupiters \citep{Cowan2011, Komacek2017}. As expected in this high-irradiation regime, the canonical model results in a thermal inversion in the atmosphere, with strong emission lines of H$_2$O and CO in the K-band spectrum. We then explored a range of chemical abundances and temperature profiles in search of models that provide a high detection significance when cross-correlated with the data. As discussed in section~\ref{sec:mol_det}, the data shows evidence for the presence of a thermal inversion in the atmosphere and emission features from CO and H$_2$O, with the CO signature being significantly stronger than H$_2$O. For the final model used in \ref{sec:mol_det}, the abundances of the prominent species are CO = 5$\times$10$^{-4}$, H$_2$O = 10$^{-5}$ and TiO = 10$^{-8}$, and the day-night energy redistribution is 10\%. The corresponding model spectrum in the observed wavelength range and the temperature profile are shown in figure \ref{fig:models}.

\begin{figure*}
\gridline{\fig{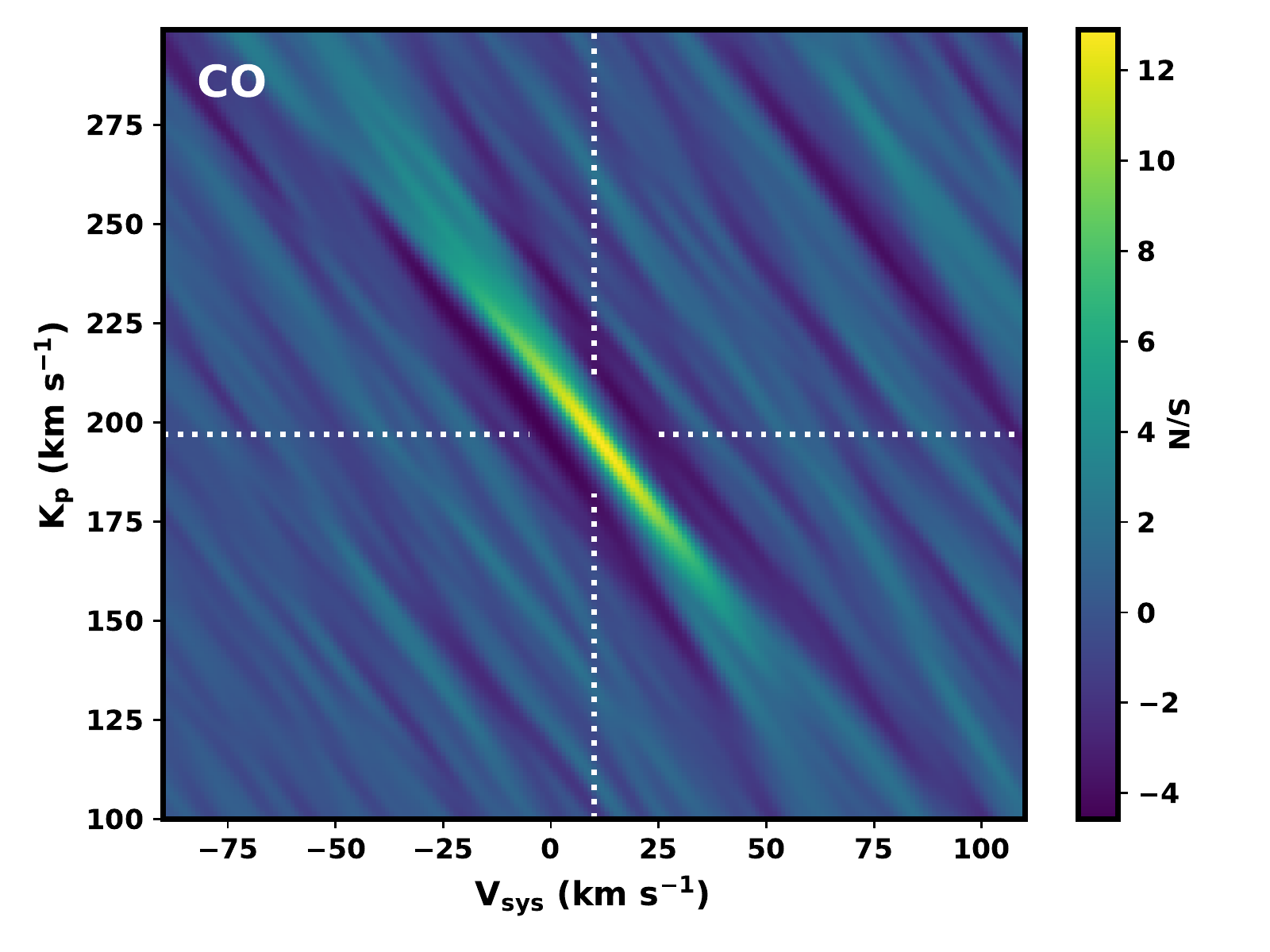}{0.48\textwidth}{}
        \fig{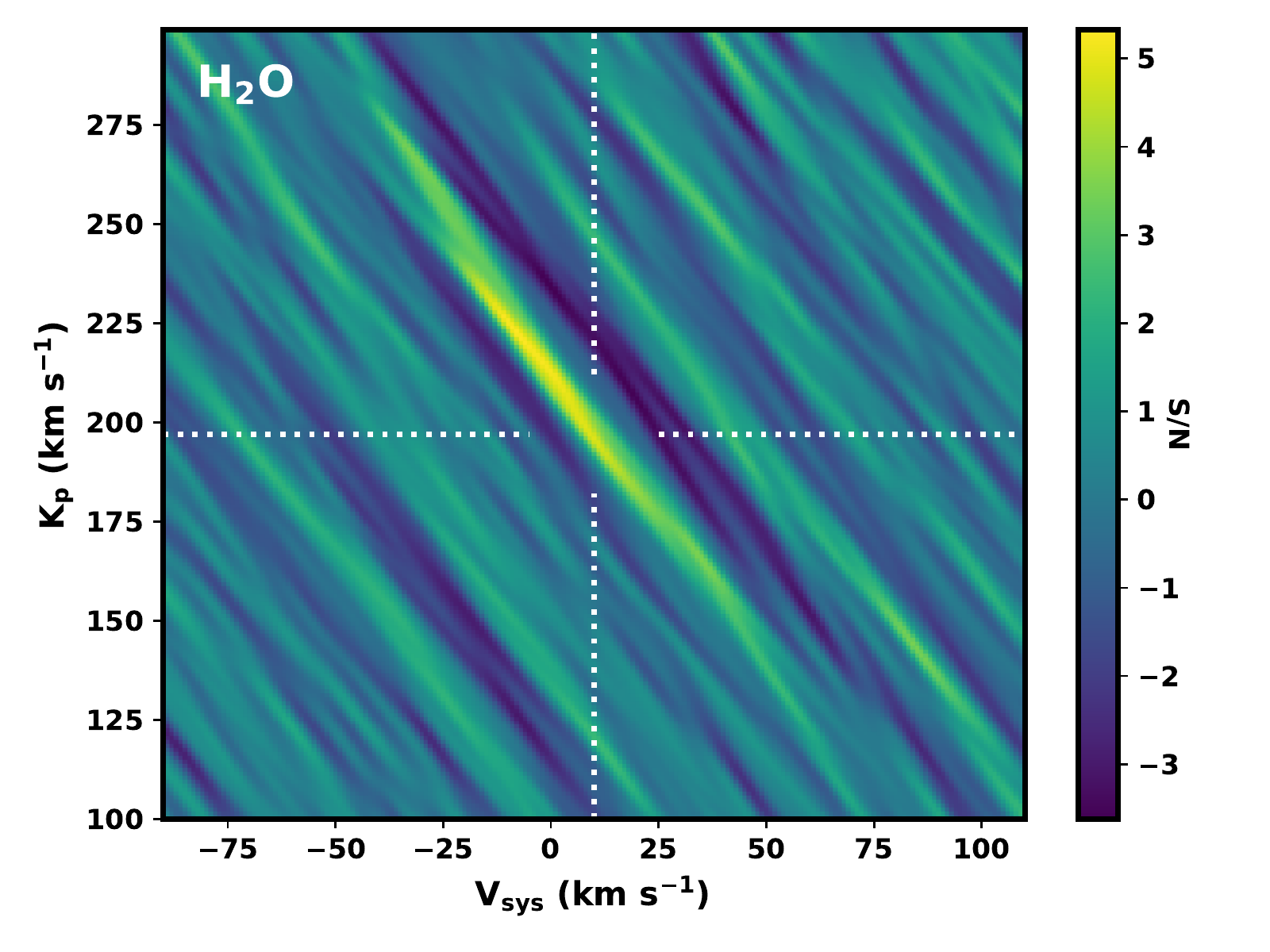}{0.48\textwidth}{}}
\caption{Cross-correlation maps for CO and H$_2$O individually in the atmosphere of MASCARA-1 b, using CRIRES+ spectra from the \pipeline pipeline. The peak significance of the detections is measured to be S/N = 12.9 and S/N = 5.3 for CO and H$_2$O, respectively. The white dotted lines indicate K$_{\mathrm{p}}$ = 197 km s$^{-1}$ and V$_{\mathrm{sys}}$ = 10 km s$^{-1}$, the signal maximum of the combined model.}
\label{fig:detections}
\end{figure*}

Before performing the cross-correlation, we remove the continuum from the model spectra. Note that this was not done to the model spectra that we injected to determine the optimal detrending. Finally, we broadened the model spectra with a Gaussian to match the resolution of CRIRES+. We use a FWHM of 2.1 and 2.7 km s$^{-1}$ when we cross-correlate with spectra from the CR2RES pipeline and the \pipeline pipeline, respectively, given the difference in resolution. As expected, the detection significance increased by broadening the models, however the difference between using a Gaussian with a FWHM of 2.1 or 2.7 km s$^{-1}$ was small.

\subsection{Cross-correlation analysis}

We cross-correlate all orders at once with a model spectrum, after sufficient telluric removal, and weight each wavelength bin according to the telluric transmission and noise level. Like \cite{cabot_detection_2020}, we define our CCF as a function of velocity and time:
\begin{equation}
    \mathrm{CCF}(v, t) = \frac{\sum_i m_i(v) w_i x_i(t)}{\sum_i m_i(v) w_i}\,,
\end{equation}
where $x_i(t)$ is the time-dependant residual spectrum at wavelength bin $i$ and $m_i(v)$ is the model spectrum Doppler shifted by velocity $v$. The weights are given by the time-averaged spectrum $T_i$ (used as a proxy for the telluric transmission) and the masking $M_i$, divided by the wavelength bin variance of the residuals $\sigma_i^2$, i.e. $w_i = M_i T_i / \sigma_i^2$. Only spectral channels that have been flagged as bad by the extraction pipeline are masked. We note that the residual variance $\sigma_i^2$, after a few singular values have been removed, is in good agreement with the variance from \eqref{eq:extract_var}.

The CCF is evaluated at velocities between -1000 to +1000 km s$^{-1}$, in steps of 1 km s$^{-1}$, where we used cubic spline interpolation to Doppler shift the model spectra. Next, we shift the CCF to a range of rest frames, considering a circular orbit with radial velocity
\begin{equation} \label{eq:orbit}
    \text{V}_\text{p}(t) = K_p \sin 2\pi \phi(t) + \Vsys - \text{V}_\text{bary}(t) \, ,
\end{equation}
where $\phi$ is the orbital phase and $\text{V}_\text{bary}$ is the time-dependant barycentric velocity correction. Thus, we make no a priori assumptions regarding the maximum radial velocity K$_{\mathrm{p}}$. We sample K$_{\mathrm{p}}$ from 100 km s$^{-1}$ to 300 km s$^{-1}$ and V$_{\mathrm{sys}}$ from -300 km s$^{-1}$ to 300 km s$^{-1}$, in steps of 1 km s$^{-1}$. The orbital parameters that we used are outlined in table \ref{tab:parameters}.

We convert the cross-correlation value of the peak to a detection significance using the S/N metric, i.e. the signal as compared to the cross-correlation noise. The noise is taken to be the standard deviation of the cross-correlation values that are more than 100 km s$^{-1}$ away from the expected $\Vsys$. For robustness, we also evaluate the S/N where the CCF standard deviation is calculated per K$_{\mathrm{p}}$ value, which is done in some studies \citep[see e.g.][]{spring_black_2022}, and further vary the V$_{\mathrm{sys}}$ range over which we calculate the standard deviation (from $\pm 200$ to $\pm 600$ km s$^{-1}$).

\subsection{Molecular detections}

\label{sec:mol_det}
\begin{figure}
\gridline{\fig{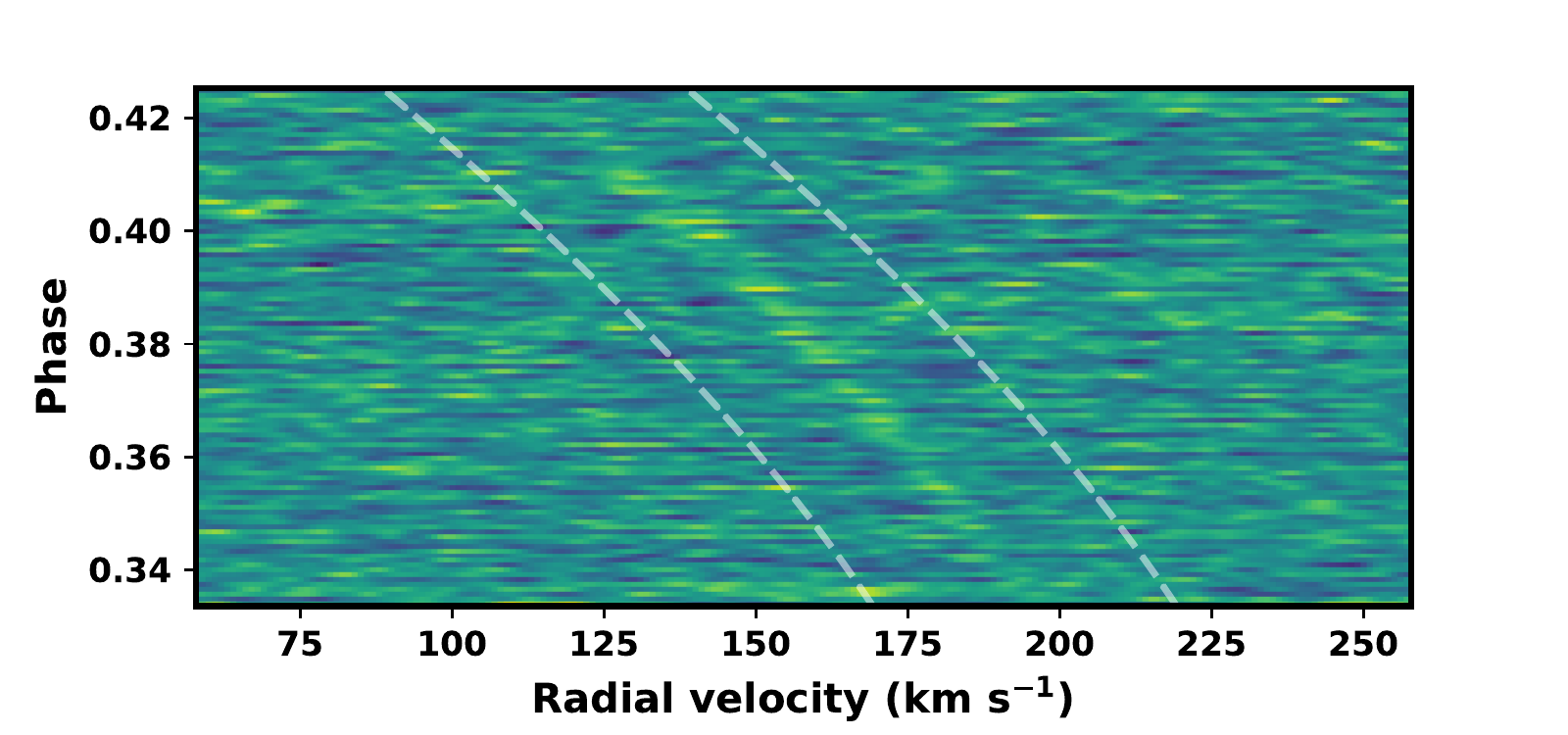}{0.48\textwidth}{}}
\vspace{-7mm}
\gridline{\fig{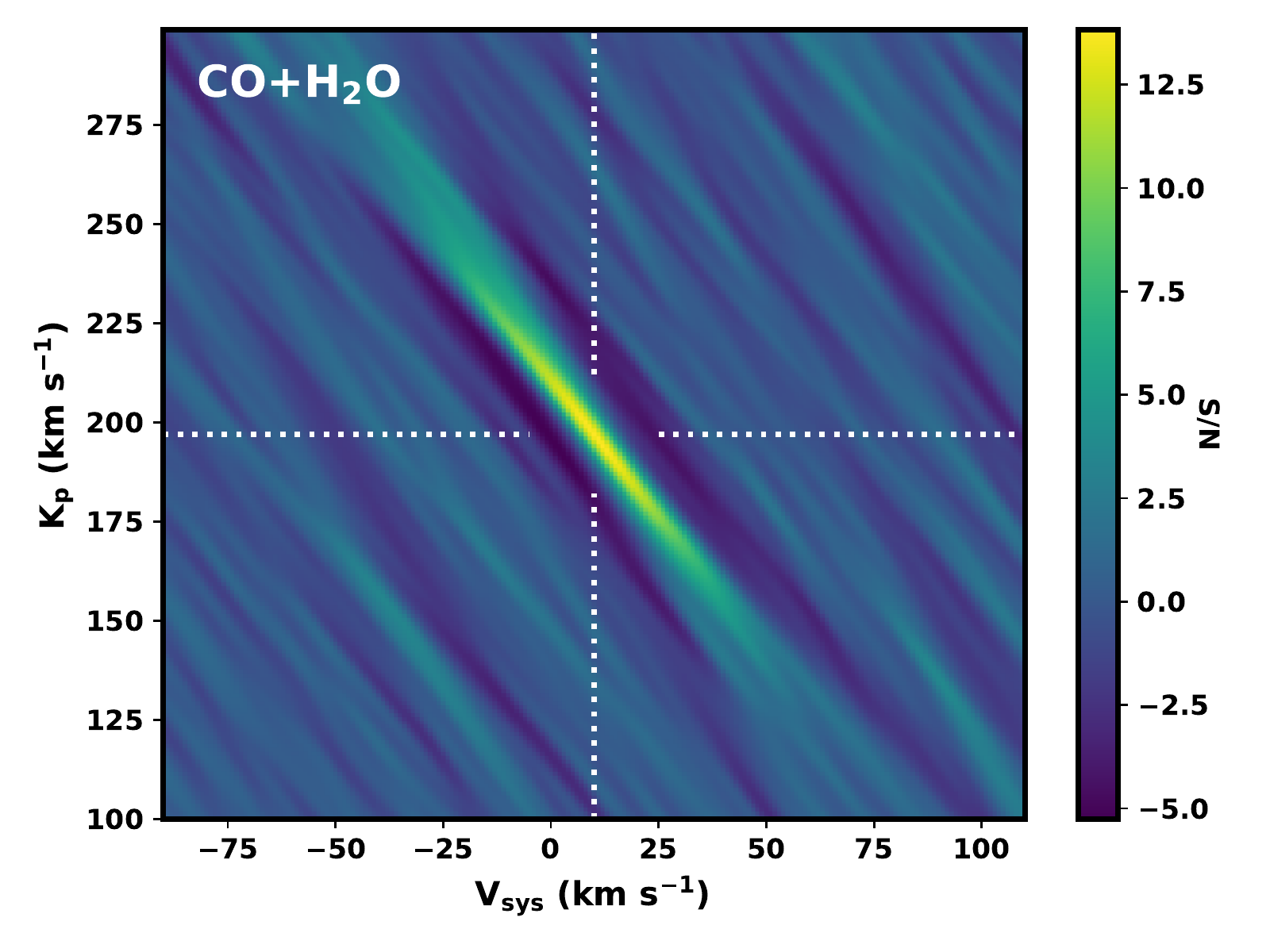}{0.48\textwidth}{}}
\caption{Detection of combined CO and H$_2$O in emission in the atmosphere of MASCARA-1 b. The top panel shows the cross-correlation function as a function of orbital phase and planet radial velocity. The white dotted lines indicate the planet radial velocity using the  $\Kp$-$\Vsys$ corresponding to the peak in the CCF map below. The bottom panel shows the S/N detection map, resulting in a significance of S/N = 13.8. White dashed lines show the $\Kp$-$\Vsys$ at the signal maximum. The atmospheric model used for the cross-correlation has combined contributions from CO and H$_2$O along with a thermal inversion as discussed in section~\ref{sec:model}.}
\label{fig:combined_detection}
\end{figure}

We find strong signatures of CO and H$_2$O emission lines in the dayside atmosphere of MASCARA-1 b. As shown in figure \ref{fig:detections}, we detect CO at S/N = 12.9 and H$_2$O at S/N = 5.3. Both these detections are made by cross-correlating a high-resolution model spectrum of MASCARA-1 b, with a thermal inversion in the atmosphere giving rise to molecular emission features, as shown in figure~ \ref{fig:models} and discussed in section~\ref{sec:model}. The maximum cross-correlation for CO is found at K$_{\mathrm{p}}$ = 193$_{-8}^{+11}$ km s$^{-1}$ and V$_{\mathrm{sys}}$ = 13$_{-8}^{+6}$ km s$^{-1}$, while the maximum for H$_2$O is at K$_{\mathrm{p}}$ = 224$_{-35}^{+11}$ km s$^{-1}$ and V$_{\mathrm{sys}}$ = -9$_{-8}^{+24}$ km s$^{-1}$. Uncertainties correspond to the S/N$_{\mathrm{peak}}-1$ contour surrounding the maximum S/N. Our estimate of V$_{\mathrm{sys}}$ is consistent with the values reported by \cite{talens_mascara-1_2017} of 8.52 $\pm$ 0.02 or 11.20 $\pm$ 0.08 km s$^{-1}$, which they obtain using two different datasets. We find that the RV semi-amplitude is also consistent with 217 $\pm$ 25 km s$^{-1}$ as calculated using parameters from \cite{talens_mascara-1_2017}. By fixing V$_{\mathrm{sys}}$ to 8.52 km s$^{-1}$ and 11.20 km s$^{-1}$, we obtain K$_{\mathrm{p}}$ = $198.5 \pm 1.8$ and $195.8 \pm 1.7$ km s$^{-1}$, respectively, using the CO signal. Likewise, using the H$_2$O signal, we obtain K$_{\mathrm{p}}$ = $197.8 \pm 3.5$ and $194.9 \pm 3.0$ km s$^{-1}$, respectively. These values are somewhat different from 204.2 $\pm$ 0.2 km s$^{-1}$, as calculated using parameters from \cite{hooton_spi-ops_2022}.

Our detection of emission features of CO and H$_2$O require the presence of a temperature inversion in the dayside atmosphere of MASCARA-1 b. This follows from the expectation that an atmosphere with temperature decreasing outward causes absorption features in the emergent spectrum whereas temperatures increasing outward, i.e. a temperature inversion, causes emission features \citep[e.g.][]{hubeny_possible_2003,fortney_unified_2008,gandhi_new_2019}. The model temperature structure with a temperature inversion is shown in figure~\ref{fig:models}. While the present results are the first detections of CO and H$_2$O emission lines in high-resolution in an exoplanet, our detection of a thermal inversion in MASCARA-1 b follows detections of thermal inversions in other UHJs, e.g. WASP-18 b \citep{sheppard_evidence_2017, arcangeli_h_2018} and WASP-33 b  \citep{haynes_spectroscopic_2015, nugroho_high-resolution_2017, nugroho_detection_2020}. In the present case, the thermal inversion in the model is caused by the strong irradiation received by this ultra-hot jupiter and the presence of visible opacity due to TiO, Na, and K. Follow-up observations in the visible could be used to search for strong visible opacities to determine the source of the temperature inversion in MASCARA-1 b.

We explored a range of self-consistent models and found that a sub-solar H$_2$O abundance was preferred. A model spectrum with similar H$_2$O and CO mixing ratios, assuming solar elemental composition, gave a lower detection significance compared to a model in which H$_2$O was depleted relative to CO. By using H$_2$O = 10$^{-5}$ and CO = 5$\times$10$^{-4}$ we obtain S/N = 13.8, with K$_{\mathrm{p}}$ = 197$_{-11}^{+10}$ km s$^{-1}$ and V$_{\mathrm{sys}}$ = 10$_{-7}^{+8}$ km s$^{-1}$. The result of the cross-correlation analysis using this atmospheric model, including both H$_2$O and CO, is shown in \ref{fig:combined_detection}. Future studies can explore constraints on the atmospheric abundances and temperature structure in more detail.

A sub-solar H$_2$O abundance in this UHJ may indicate that H$_2$O is partially dissociated as a result of the high dayside temperature of the planet. The dissociation of H$_2$O in UHJs has been investigated in several recent studies \citep[e.g.][]{arcangeli_h_2018,parmentier_thermal_2018, lothringer_extremely_2018}. Such a scenario would result in reduced H$_2$O spectral features in the infrared as inferred here, which can also be constrained with lower-resolution infrared spectra from the Hubble Space Telescope (HST) and the James Webb Space Telescope (JWST). Furthermore, strong visible opacity resulting from the formation of H$^-$ could act to increase the strength and probability of a temperature inversion \citep{arcangeli_h_2018, lothringer_extremely_2018, parmentier_thermal_2018}. Recent high-resolution near-infrared ($\sim$1-1.7 $\mu$m) observations of the UHJs WASP-33 b \citep{nugroho_first_2021} and WASP-76 b \citep{landman_detection_2021} found evidence for thermal dissociation of H$_2$O by detected hydroxyl radical (OH). Such observations could help investigate if H$_2$O is dissociated in the atmosphere of MASCARA-1 b. If the H$_2$O depletion is found to be not due to dissociation, then another possibility would be a high C/O ratio in the atmosphere \citep[][]{madhu2012,moses2013}.

Depending on how we choose to calculate the CCF noise, as described above, the S/N of the CO detection varies from 11.6 - 13.8 and the H$_2$O detection varies from 4.8 - 5.4, hence this should be regarded at the uncertainty of the detection S/N. Further note that the negative cross-correlation wings surrounding the peak of the CO and CO + H$_2$O detections are not indicative of the real average line shape. We see the same negative wings in the injected signal, where the effect gets worse the more singular values we remove. Similar anti-correlation wings can be seen in other high S/N detections that use PCA/SYSREM detrending \citep[e.g.][]{line_solar_2021}. This effect is discussed in \cite{spring_black_2022}.

Moreover, we are now in a position to evaluate the effects of the data reduction on the final detection significance. In table \ref{tab:CO_significance} of Appendix \ref{sec:appendix} we provide the S/N of the CO detection using spectra reduced using both the \pipeline pipeline and the CR2RES pipeline. We find that the maximum S/N is achieved with the \pipeline pipeline.

\section{Summary and discussion} \label{sec:discussion}

We present the first results with CRIRES+ using observations from the science verification run. We analyse data of four exoplanet observations in the K-band to provide early insights into the performance of the instrument. To do this, we employ two reduction pipelines: the official ESO CR2RES pipeline and our custom-built \pipeline pipeline. In every step of the analysis, we compare the results from the two pipelines. We assess the performance of CRIRES+ using two key metrics, the spectral resolution and the S/N, obtained across the wide range of targets and observing conditions. 

We find that CRIRES+ is capable of resolutions of $R \gtrsim$100,000 when using the adaptive optics during good observing conditions.
We measure the resolving power by fitting a high-resolution telluric model to the spectra and find that the resolution well exceeds the advertised minimum resolution of $R = 80,000$. We show that the high resolution obtained is a result of the stellar PSF being significantly narrower than the width of the 0.2" slit for very good observing conditions with adaptive optics. This shows that the FWHM of the star in the spatial direction along the slit can be used as a quick proxy for the spectral resolution, even before extraction, at least when averaging over multiple orders and when the stellar PSF is not limited by the slit. We also find that seeing impacts the adaptive optics performance, which in turn affects the resolution. This is well demonstrated by comparing the data of MASCARA-1 b, observed during excellent seeing conditions, to that of WASP-20 b and LTT 9779 b, which were observed during worse conditions; resulting in a wider stellar PSF in the latter cases. Consequentially, this is reflected in the spectral resolution of these observations, as shown in table \ref{tab:metrics}.

The fluxes of the spectra extracted using the two pipelines, CR2RES and ExoRES, are in good agreement. The S/N of the observations reduced with the \pipeline pipeline agree with both the empirical S/N metric and expectations from the ESO CRIRES+ ETC for the same observing conditions, as shown in figure  \ref{fig:SNR_spectra}. Overall, our results demonstrate that CRIRES+ is performing as expected.

Finally, we demonstrate the on-sky capabilities of CRIRES+ in the context of atmospheric characterisation of exoplanets using a case study of the ultra-hot Jupiter MASCARA-1 b. We detect the presence of CO and H$_2$O in the dayside atmosphere with high significance. CO and H$_2$O are found in emission with a significance of S/N = 12.9 (K$_{\mathrm{p}}$ = 193$_{-8}^{+11}$ km s$^{-1}$ and V$_{\mathrm{sys}}$ = 13$_{-8}^{+6}$ km s$^{-1}$) and S/N = 5.3 (K$_{\mathrm{p}}$ = 224$_{-35}^{+11}$ km s$^{-1}$ and V$_{\mathrm{sys}}$ = -9$_{-8}^{+24}$ km s$^{-1}$), respectively. Using a model with both CO and H$_2$O, we obtain an increased significance of S/N = 13.8 (K$_{\mathrm{p}}$ = 197$_{-11}^{+10}$ km s$^{-1}$ and V$_{\mathrm{sys}}$ = 10$_{-7}^{+8}$ km s$^{-1}$). Through the emission features of CO and H$_2$O, we also detect a temperature inversion in the dayside atmosphere of MASCARA-1 b. Both the systemic velocity and the RV semi-amplitude are consistent with the orbital parameters provided by \cite{talens_mascara-1_2017}.

We perform a range of robustness checks to make sure the significance of our detections are not biased. When performing the detrending, as outlined in section \ref{sec:detrending}, we select the optimal number of singular values to remove using a method that is insensitive to the $\Kp$ and $\Vsys$ of interest. We achieve this by performing an injection test that does not depend on the noise at the site of injection. To limit overfitting further, we perform the same detrending on all orders. The fact that the injection test shows that different models, injected at a range of different $\Kp$ and $\Vsys$, prefer very similar levels of detrending, indicate that the parameters determining the optimal detrending may be more universal than previously thought. 

Our current findings for MASCARA-1 b suggest that the H$_2$O abundance might be significantly lower compared to CO. This is evident from the weaker H$_2$O signal compared to the CO signal, despite the presence of adequate spectral features in the observed band. Furthermore, when cross-correlating a combined model with both CO and H$_2$O present, a H$_2$O-depleted model was strongly favoured over one where CO and H$_2$O are similar in abundance, as  may be expected in thermochemical equilibrium with solar elemental abundances without thermal dissociation. More detailed abundance estimates in the future may be able to determine if the potentially low H$_2$O abundance is due to thermal dissociation of H$_2$O, as has been suggested to be possible for ultra-hot Jupiters.

Overall, our first look at the performance metrics of CRIRES+ reveals a promising outlook for atmospheric characterisation of exoplanets. Under optimal observing conditions we find that CRIRES+ is performing as expected or better. This opens the avenue for high-precision and high-resolution atmospheric characterisation of diverse exoplanets, from hot Jupiters to cooler and smaller exoplanets. This is a particularly opportune moment considering the complementarity that high-resolution infrared spectroscopy of exoplanets from the ground will offer to space-based transit spectroscopy with JWST, heralding a new era in exoplanet science.

\vspace{7mm}

This work is supported by research grants to N.M. from the MERAC Foundation, Switzerland, and the UK Science and Technology Facilities Council (STFC) Center for Doctoral Training (CDT) in Data intensive science at the University of Cambridge (STFC grant number ST/P006787/1). N.M. and M.H. acknowledge support from these sources towards the doctoral studies of M.H. N.M. thanks Siddharth Gandhi and Anjali Piette for helpful discussions. This work is based on observations collected at the European Southern Observatory (ESO) under programmes: 107.22SX.001, 107.22TQ.001, 107.22TE.001, and 107.22U7.001, using the CRIRES+ spectrograph on the ESO Very Large Telescope (VLT). We thank the teams associated with the above programs for preparing the observations used in this work. We thank ESO for providing the data through the ESO Science Archive.

\bibliographystyle{aasjournal} 
\bibliography{references, refs_extra}
\appendix

\section{Detection significance comparison} \label{sec:appendix}
We compare the detection significance of CO in MASCARA-1 b using spectra reduced
using different extractions. The result is presented in table \ref{tab:CO_significance}. We find that the highest significance detection is obtained by the \pipeline pipeline, which also required the lowest amount of detrending. Moreover, we note that the detection S/N is slightly lower for the spectra extracted using a swath width of 1600 pixels, as compared to using a smaller swath width.

\begin{table*}[h]
\caption{Detection significance of CO in MASCARA-1 b for different extraction methods. All CR2RES pipeline instances used an oversampling factor of 12. The extraction swath width is given in pixels.}
\movetableright=1.15in
\begin{tabular}{lllll}
\hline
Pipeline & Swath width & Band subtraction & Optimal $k$ & S/N \\ \hline
\pipeline    &  - &  yes  & 7 & 12.9 \\
CR2RES    & 400 & no & 11 & 11.0 \\
CR2RES    & 800 & no & 9 & 11.2 \\
CR2RES    & 1600 & no & 9 & 10.5 \\
CR2RES    & 400 & yes & 14 & 11.1 \\
CR2RES    & 800 & yes & 12 & 10.9 \\
CR2RES    & 1600 & yes & 15 & 9.8 \\ \hline
\end{tabular}
\label{tab:CO_significance}
\end{table*}

\label{lastpage}
\end{document}